\documentclass[12pt]{article}
\usepackage{amsfonts,graphicx,amsmath,amsthm,amssymb,epsfig}
\usepackage{float}
\allowdisplaybreaks

\begin{document}

\title{\bf Effects of Charge and Gravitational Decoupling on Complexity and Isotropization of Anisotropic Models}
\author{M. Sharif$^1$ \thanks{msharif.math@pu.edu.pk} and Tayyab Naseer$^{1,2}$ \thanks{tayyabnaseer48@yahoo.com}\\
$^1$ Department of Mathematics and Statistics, The University of Lahore,\\
1-KM Defence Road Lahore-54000, Pakistan.\\
$^2$ Department of Mathematics, University of the Punjab,\\
Quaid-i-Azam Campus, Lahore-54590, Pakistan.}

\date{}
\maketitle

\begin{abstract}
This paper constructs two immediate extensions of the existing
anisotropic solutions in the context of Einstein-Maxwell framework
by employing minimal geometric deformation. To achieve this, we
assume a static spherical interior initially filled with anisotropic
fluid and call it a seed source. We extend this matter configuration
by including a new source whose impact on the self-gravitating
system is governed by a decoupling parameter. The charged field
equations analogous to the total fluid source are formulated. We
then implement a transformation on the radial metric potential that
divides the field equations into two new under-determined systems,
corresponding to the initial and new sources. The first of them is
addressed by taking two well-known metric ansatz so that the number
of unknowns can be tackled. Further, the vanishing of total
anisotropy and complexity-free constraints are used to solve the
second set. The estimated radius and mass of a compact star
$4U~1820-30$ is utilized to interpret the resulting solutions
graphically for different values of the charge and decoupling
parameter $\omega$. We conclude that our both developed models are
physically acceptable for all parametric values except $\omega=1$.
\end{abstract}
{\bf Keywords:} Geometric deformation; Anisotropy; Self-gravitating system. \\
{\bf PACS:} 04.20.Jb; 95.36.+x.

\section{Introduction}

General theory of relativity ($\mathrm{GR}$) has gained significant
acceptance throughout the scientific community as it is considered
the most desirable tool for understanding the nature of
gravitational field. This theory relates the matter configuration
with the geometry of a spacetime structure characterized by the
energy-momentum tensor ($\mathrm{EMT}$) and the Einstein tensor,
respectively through Einstein's field equations. The exact/numerical
formulation of their solutions has become an interesting topic among
physicists through which they study the structural properties of
self-gravitating systems. Schwarzschild pioneered this work by
calculating the analytical solution of such equations that prompted
researchers to extend this in the context of $\mathrm{GR}$ and other
modified theories. For the first time in $1916$, he started with the
assumption of a spherical geometry possessing uniform density and
formulated the corresponding exterior \cite{1} and interior
solutions \cite{2}. The literature offers a large number of
solutions in recent times describing geometrical structures coupled
with different fluid distributions.

Among all solutions of the field equations, those which can be used
to model physically acceptable compact bodies have made remarkable
achievements in the literature. However, it is rather difficult to
calculate such solutions because the field equations involve
multiple geometric terms that make them highly non-linear. Such
non-linearity forced researchers to develop certain approaches which
can be used to solve these equations and produce physically relevant
results. The novel and innovative strategy among them is the
gravitational decoupling which is empowered to tackle with several
factors reigning the interior distribution such as anisotropy,
dissipation flux, expansion scalar and shear, etc. This technique
leads the field equations in a new reference frame where each fluid
source is characterized by an individual system of equations, and
hence, each set can be solved easily. Further, decoupling offers two
different types, namely minimal ($\mathrm{MGD}$) and extended
geometric deformations, transforming radial and both temporal/radial
metric potentials, respectively.

Ovalle \cite{29} recently pioneered the $\mathrm{MGD}$ scheme in the
braneworld ($\mathrm{BW}$) scenario and discussed compact objects
characterized by exact solutions. Following this, Ovalle and Linares
\cite{30} assumed Tolman IV spacetime as the isotropic solution and
developed its corresponding anisotropic analog in the context of
$\mathrm{BW}$ through the same technique. The study of different
phenomena have been made possible through this technique, such as
the derivation of solutions corresponding to spherical \cite{3}
interior, originating acceptable inner fluid solution \cite{4,4a},
studying microscopic properties of black holes \cite{5,6} and
discussing the impact of Weyl stresses \cite{7}. The perturbation
was applied to the field equations in $\mathrm{BW}$ and bulk, and
results were found compatible with each other \cite{8}-\cite{12}.
The $\mathrm{MGD}$ technique was then extended by transforming both
$g_{tt}$ and $g_{rr}$ metric coefficients of a static sphere to
obtain a modified form of the Schwarzschild geometry in
$\mathrm{BW}$ \cite{31}.

Initially, a self-gravitating geometry was considered to be coupled
with the isotropic fluid in which the pressure acts the same in each
direction, usually radial and tangential. However, the pioneering
work by Jeans made some revolutions in the literature, suggesting
that anisotropy may occur in the interior due to the presence of
multiple factors \cite{13}. A theoretical analysis has been done by
Ruderman \cite{14} from which he found that the heavily stellar
models (i.e., with energy density not less than $10^{15}~g/cm^3$)
may possess anisotropy. A compact star surrounded by a high magnetic
field \cite{15}-\cite{18} as well as certain other elements
\cite{19}-\cite{21} must contain anisotropy in its interior. Ovalle
et al. \cite{33} considered that a spherical fluid and Schwarzschild
vacuum spacetime do not exchange energy and momentum with each
other, and developed a new anisotropic solution by using
$\mathrm{MGD}$ approach. Anisotropic extensions to different
isotropic models such as Heintzmann, Duragpal-Fuloria and Tolman VII
spacetimes have also been obtained \cite{36a}-\cite{37a}. Sharif and
his collaborators \cite{35,35a} have done the same analysis for
charged/uncharged spacetimes in $\mathrm{GR}$ as well as different
extended theories and found acceptable results. We have developed
physically relevant anisotropic compact models in a non-minimally
coupled gravity for certain values of the corresponding parameters
\cite{37f}.

The inclusion of the electric charge has a significant role in the
study of compact structures. Electromagnetic forces are considered
as the most significant ingredients that help to to reduce the
attractive force of gravity. Therefore, a sufficiently enough amount
of the charge is needed to resist the gravitational force and hence,
the stability of a star is preserved. Bekenstein \cite{37fa}
considered a dynamical spherical model coupled with the
electromagnetic field and deduced that the force produced by the
charge has repulsive behavior, helping the geometry to maintain its
stability. This work has been extended by Esculpi and Aloma
\cite{37fb} for the case of anisotropic fluid from which they
observed that both positive anisotropy and charge have repulsive
effects. The structure of anisotropic charged stars has been studied
by taking a power-law interior charge distribution in terms of
exterior charge and radius of the considered compact object, and
physically relevant results were obtained \cite{42aaa,37fc}. Maurya
et al. \cite{37fd} analyzed the interior distribution of charged
stellar models configured with the baryonic fluid via the Karmarkar
condition. They solved the corresponding Einstein-Maxwell field
equations and observed the resulting matter variables to be
dependent on the electromagnetic field.

Herrera \cite{37g} recently developed a new definition of the
complexity for a static sphere that encounters the factors which
complicate the system. Initially, Bel decomposed the curvature
tensor into its orthogonal components and found certain scalars,
named the structure scalars. Herrera adopted the same formulation
for anisotropic fluid to obtain the corresponding scalars that were
appeared to be associated with different physical parameters. One of
those scalars possesses both anisotropy and inhomogeneous density,
came up with his criteria and thus named as the complexity factor.
Such a definition has been observed highly acceptable in every field
of science. Herrera et al. \cite{37h} used this definition to
discuss the evolutionary patterns for dynamical dissipative
geometry. Yousaf et al. \cite{37i,37ia} performed the same analysis
in a particular non-minimal theory and examined the impact of
modified gravity on charged/uncharged spheres and cylinders. The
complexity-free models can be obtained once we substitute the
corresponding complexity factor equals to zero. This constraint has
been employed to static self-gravitating spheres, and several
physically relevant objects are modeled \cite{37j}-\cite{37la}.

This article extends the anisotropic spherical solutions to the
Einstein-Maxwell framework through the $\mathrm{MGD}$ technique. To
do this work, we follow the structure defined in the succeeding
lines. Section \textbf{2} introduces the influence of charge in
Einstein's field equations for a static spherical interior
possessing two sources, i.e., initial anisotropic and newly added
fluids. A well-known geometric transformation is applied on the
formulated field equations in section \textbf{3}, resulting into two
sets. Different constraints on both systems of equations are
employed independently and find the corresponding solutions in
sections \textbf{4} and \textbf{5}. Section \textbf{6} discusses the
influence of the decoupling technique and charge on physical
attributes of the resulting models. On a final note, we summarize
all our results in section \textbf{7}.

\section{Static Spherical Geometry and Einstein-Maxwell Framework}

We present the field equations in this section that describe a
static spherical spacetime influenced by an electromagnetic field.
The spherical geometry $(t,r,\theta,\vartheta)$ is considered to
split into two sectors, namely exterior and interior regions over
the hypersurface $\Sigma$. In order to perform our analysis, the
interior spacetime is given by the line element as
\begin{equation}\label{g6}
ds^2=-e^{\xi_1} dt^2+e^{\xi_2} dr^2+r^2 d\Omega^2,
\end{equation}
where $\xi_1=\xi_1(r),~\xi_2=\xi_2(r)$ and
$d\Omega^2=d\theta^2+\sin^2\theta d\vartheta^2$. The presence of
Lagrangian corresponding to the additional source and
electromagnetic field in the Einstein-Hilbert action produces the
field equations given by
\begin{equation}\label{g2}
\mathrm{G}_{\lambda\chi}=\mathrm{R}_{\lambda\chi}-\frac{1}{2}\mathrm{R}g_{\lambda\chi}=8\pi\bar{\mathrm{T}}_{\lambda\chi},
\end{equation}
with
\begin{equation}\label{g1}
\bar{\mathrm{T}}_{\lambda\chi}=\mathrm{T}_{\lambda\chi}+\mathrm{E}_{\lambda\chi}+\omega\mathrm{Z}_{\lambda\chi},
\end{equation}
where the quantities are expressed as
\begin{itemize}
\item $\mathrm{G}_{\lambda\chi}$ is the Einstein tensor that
characterizes the geometry,
\item $\mathrm{R}_{\lambda\chi},~\mathrm{R}$ and $g_{\lambda\chi}$
indicate the Ricci tensor, Ricci scalar and the metric tensor,
respectively,
\item $\mathrm{T}_{\lambda\chi}$ is the usual matter $\mathrm{EMT}$,
\item $\mathrm{E}_{\lambda\chi}$ is the electromagnetic $\mathrm{EMT}$,
\item $\mathrm{Z}_{\lambda\chi}$ is the additional matter source gravitationally
associated with $\mathrm{T}_{\lambda\chi}$.
\end{itemize}
Since our aim is to study the complexity of a self-gravitating
system, we must assume anisotropy in the interior initially whose
$\mathrm{EMT}$ is defined by
\begin{equation}\label{g5}
\mathrm{T}_{\lambda\chi}=(P_t+\rho)\mathrm{W}_{\lambda}\mathrm{W}_{\chi}+P_t
g_{\lambda\chi}-\left(P_t-P_r\right)\mathrm{X}_\lambda\mathrm{X}_\chi,
\end{equation}
where $P_r,~P_t,~\rho,~\mathrm{W}_{\chi}$ and $\mathrm{X}_{\chi}$
are the radial/tangential pressure elements, energy density,
four-velocity and the four-vector, respectively. We consider a
co-moving frame of reference that gives rise to the following
quantities in accordance with Eqs.\eqref{g6} and \eqref{g5} as
\begin{equation}\nonumber
\mathrm{X}^\chi=(0,e^{\frac{-\xi_2}{2}},0,0), \quad
\mathrm{W}^\chi=(e^{\frac{-\xi_1}{2}},0,0,0),
\end{equation}
fulfilling the relations
$$\mathrm{X}^\chi \mathrm{W}_{\chi}=0, \quad \mathrm{X}^\chi
\mathrm{X}_{\chi}=1, \quad \mathrm{W}^\chi \mathrm{W}_{\chi}=-1.$$

The electromagnetic field characterized by the $\mathrm{EMT}$ is
expressed as
\begin{equation}\label{g1}
\mathrm{E}_{\lambda\chi}=-\frac{1}{4\pi}\left[\frac{1}{4}g_{\lambda\chi}\mathrm{F}^{\zeta\beta}\mathrm{F}_{\zeta\beta}
-\mathrm{F}^{\zeta}_{\chi}\mathrm{F}_{\lambda\zeta}\right].
\end{equation}
Here, the Maxwell field tensor is represented by
$\mathrm{F}_{\zeta\beta}=\Psi_{\beta;\zeta}-\Psi_{\zeta;\beta}$ with
the four-potential $\Psi_{\zeta}=\Psi(r)\delta^{0}_{\zeta}$. The
Maxwell equations can concisely be written as follows
\begin{equation*}
\mathrm{F}^{\zeta\beta}_{;\beta}=4\pi \jmath^{\zeta}, \quad
\mathrm{F}_{[\zeta\beta;\nu]}=0,
\end{equation*}
with $\jmath^{\zeta}=\varpi \mathrm{W}^{\zeta}$ and $\varpi$ being
the current and charge densities, respectively. The left side of the
above equations yields in this framework as
\begin{equation}\nonumber
\Psi''+\frac{1}{2r}\big\{4-r(\xi_1'+\xi_2')\big\}\Psi'=4\pi\varpi
e^{\frac{\xi_1}{2}+\xi_2},
\end{equation}
where $'=\frac{\partial}{\partial r}$. Integrating the above
equation, we have
\begin{equation}\nonumber
\Psi'=\frac{\mathrm{q}}{r^2}e^{\frac{\xi_1+\xi_2}{2}},
\end{equation}
with $\mathrm{q}=\int_0^r \varpi
e^{\frac{\xi_2}{2}}\bar{y}^2d\bar{y}$ being the total charge in the
interior. The non-vanishing components of the $\mathrm{EMT}s$
\eqref{g5} and \eqref{g1} are
\begin{eqnarray}\nonumber
&\mathrm{T}_{00}=\rho e^{\xi_1},\quad
\mathrm{T}_{11}=P_re^{\xi_2},\quad
\mathrm{T}_{22}=P_tr^2=\frac{\mathrm{T}_{33}}{\sin^2\theta},\\\nonumber
&\mathrm{E}_{00}=\frac{\mathrm{q}^2e^{\xi_1}}{8\pi r^4}, \quad
\mathrm{E}_{11}=-\frac{\mathrm{q}^2e^{\xi_2}}{8\pi r^4}, \quad
\mathrm{E}_{22}=\frac{\mathrm{q}^2}{8\pi
r^2}=\frac{\mathrm{E}_{33}}{\sin^2\theta}.
\end{eqnarray}

The three independent field equations corresponding to a sphere
\eqref{g6} are obtained from Eq.\eqref{g2} as
\begin{align}\label{g8}
8\pi\rho+\frac{\mathrm{q}^2}{r^4}-8\pi\omega\mathrm{Z}^{0}_{0}
&=\frac{1}{r^2}-e^{-\xi_2}\left(\frac{1}{r^2}-\frac{\xi_2'}{r}\right),\\\label{g9}
8\pi{P}_r-\frac{\mathrm{q}^2}{r^4}+8\pi\omega\mathrm{Z}^{1}_{1}
&=e^{-\xi_2}\left(\frac{1}{r^2}+\frac{\xi_1'}{r}\right)-\frac{1}{r^2},\\\label{g10}
8\pi{P}_t+\frac{\mathrm{q}^2}{r^4}+8\pi\omega\mathrm{Z}^{2}_{2}
&=\frac{e^{-\xi_2}}{4}\left[\xi_1'^2-\xi_1'\xi_2'+2\xi_1''-\frac{2\xi_2'}{r}
+\frac{2\xi_1}{r}\right].
\end{align}
The conservation equation in this case can be obtained by taking the
covariant divergence of the total fluid (seed, charge and
additional), i.e.,
$\nabla^\chi(\mathrm{T}_{\lambda\chi}+\mathrm{E}_{\lambda\chi}+\omega\mathrm{Z}_{\lambda\chi})=0$
as
\begin{align}\nonumber
&\frac{dP_r}{dr}+\frac{\xi_1'}{2}\big(\rho+P_r\big)+\frac{\omega\xi_1'}{2}
\big(\mathrm{Z}_{1}^{1}-\mathrm{Z}_{0}^{0}\big)+\frac{2}{r}\big(P_r-P_t\big)\\\label{g12}
&+\omega\frac{d\mathrm{Z}_{1}^{1}}{dr}+\frac{2\omega}{r}\big(\mathrm{Z}_{1}^{1}
-\mathrm{Z}_{2}^{2}\big)-\frac{\mathrm{qq}'}{4\pi r^4}=0.
\end{align}
The above condition must hold for the system to be in a stable
equilibrium, called the Tolman-Opphenheimer-Volkoff (TOV) equation.
The mass function can be specified in terms of geometry as well as
matter distribution. The geometric definition is given by
\begin{align}\label{g12a}
\mathrm{m}(r)=\frac{r}{2}\bigg(1-\frac{1}{e^{\xi_2}}+\frac{\mathrm{q}^2}{r^2}\bigg).
\end{align}

On the other hand, the mass in relation with the energy density and
charge can be obtained through Eqs.\eqref{g8} and \eqref{g12a} as
\begin{align}\nonumber
\mathrm{m}(r)&=4\pi\int_0^r\rho\bar{y}^2d\bar{y}+\int_0^r\frac{\mathrm{qq}'}{\bar{y}}d\bar{y}
+4\pi\omega\int_0^r\mathrm{Z}_{0}^{0}\bar{y}^2d\bar{y}\\\label{g12b}
&=4\pi\int_0^r\rho\bar{y}^2d\bar{y}+\frac{1}{2}\int_0^r\frac{\mathrm{q}^2}{\bar{y}^2}d\bar{y}+\frac{\mathrm{q}^2}{2r}
+4\pi\omega\int_0^r\mathrm{Z}_{0}^{0}\bar{y}^2d\bar{y},
\end{align}
where the first three terms correspond to the mass of charged fluid
distribution and the last term defines the mass related to the new
source. We use Eq.\eqref{g9} to determine the value of $\xi_1'$ in
terms of the mass function \eqref{g12a} as
\begin{align}\label{g12c}
\xi_1'=\frac{2\big\{4\pi\big(P_r+\omega\mathrm{Z}^{1}_{1}\big)r^4+\mathrm{m}r-\mathrm{q}^2\big\}}
{r\big(r^2-2\mathrm{m}r+\mathrm{q}^2\big)}.
\end{align}
Switching the above value into Eq.\eqref{g12}, we get
\begin{align}\nonumber
&\frac{dP_r}{dr}+\bigg[\frac{4\pi\big(P_r+\omega\mathrm{Z}^{1}_{1}\big)r^4+\mathrm{m}r-\mathrm{q}^2}
{r\big(r^2-2\mathrm{m}r+\mathrm{q}^2\big)}\bigg]
\big\{\rho+P_r+\omega\big(\mathrm{Z}_{1}^{1}-\mathrm{Z}_{0}^{0}\big)\big\}\\\label{g12d}
&+\frac{2}{r}\big(P_r-P_t\big)+\omega\frac{d\mathrm{Z}_{1}^{1}}{dr}+\frac{2\omega}{r}\big(\mathrm{Z}_{1}^{1}
-\mathrm{Z}_{2}^{2}\big)-\frac{\mathrm{qq}'}{4\pi r^4}=0,
\end{align}
where $\Pi=P_t-P_r$ and
$\Pi_{\mathrm{Z}}=\mathrm{Z}_{2}^{2}-\mathrm{Z}_{1}^{1}$ are the
anisotropic factors corresponding to the seed and additional fluid
sources, respectively.

\section{Gravitational Decoupling}

Since we include an additional source in the initial anisotropic
fluid, the corresponding field equations now become difficult to
solve due to the increment of unknowns, i.e.,
$(\xi_1,\xi_2,\mathrm{q},\rho,P_t,P_r,\mathrm{Z}_{0}^{0},\mathrm{Z}_{1}^{1},\mathrm{Z}_{2}^{2})$.
Thus, we need to adopt certain approach or constraints and reduce
the degrees of freedom to get an exact solution. On that note, we
start off with a systematic technique (referred to the gravitational
decoupling \cite{33}) whose implementation on the field equations
makes it possible to find their solution. The exciting feature of
this strategy is that it transforms the temporal/radial metric
potentials to a new frame of reference and make the set of equations
easy to handle. For this, we consider the following metric as a
solution to Eqs.\eqref{g8}-\eqref{g10} given by
\begin{equation}\label{g15}
ds^2=-e^{\xi_3(r)}dt^2+\frac{1}{\xi_4(r)}dr^2+r^2d\Omega^2.
\end{equation}
In this context, the metric components linearly transform as
\begin{equation}\label{g16}
\xi_3\rightarrow\xi_1=\xi_3+\omega\mathrm{f}, \quad \xi_4\rightarrow
e^{-\xi_2}=\xi_4+\omega\mathrm{t},
\end{equation}
along with the temporal $\mathrm{f}$ and radial $\mathrm{t}$
deformation functions.

We choose the $\mathrm{MGD}$ scheme, thus only the $g_{rr}$
component is deformed in the following, while keeping $g_{tt}$
potential remains unchanged, i.e., $\mathrm{t} \rightarrow
\bar{\mathrm{t}},~\mathrm{f} \rightarrow 0$. Equation \eqref{g16}
now turns into
\begin{equation}\label{g17}
\xi_3\rightarrow\xi_1=\xi_3, \quad \xi_4\rightarrow
e^{-\xi_2}=\xi_4+\omega\bar{\mathrm{t}},
\end{equation}
where $\bar{\mathrm{t}}=\bar{\mathrm{t}}(r)$. It is important to
note that the spherical symmetry remains preserved by these linear
mappings. We implement the transformation \eqref{g17} on
Eqs.\eqref{g8}-\eqref{g10} and obtain two systems. The first set
portraying the seed fluid source is acquired for $\omega=0$ as
\begin{align}\label{g18}
8\pi\rho+\frac{\mathrm{q}^2}{r^4}
&=e^{-\xi_2}\bigg(\frac{\xi_2'}{r}-\frac{1}{r^2}\bigg)+\frac{1}{r^2},\\\label{g19}
8\pi{P}_r-\frac{\mathrm{q}^2}{r^4}
&=e^{-\xi_2}\bigg(\frac{1}{r^2}+\frac{\xi_1'}{r}\bigg)-\frac{1}{r^2},\\\label{g20}
8\pi{P}_t+\frac{\mathrm{q}^2}{r^4}
&=\frac{e^{-\xi_2}}{4}\bigg(\xi_1'^2-\xi_2'\xi_1'+2\xi_1''-\frac{2\xi_2'}{r}+\frac{2\xi_1'}{r}\bigg).
\end{align}
Furthermore, the impact of the newly added source
$\mathrm{Z}_{\lambda\chi}$ is encoded by the following set and can
obtained for $\omega=1$ as
\begin{align}\label{g21}
&8\pi\mathrm{Z}_{0}^{0}=\frac{1}{r}\bigg(\bar{\mathrm{t}}'+\frac{\bar{\mathrm{t}}}{r}\bigg),\\\label{g22}
&8\pi\mathrm{Z}_{1}^{1}=\frac{\bar{\mathrm{t}}}{r}\bigg(\xi_1'+\frac{1}{r}\bigg),\\\label{g23}
&8\pi\mathrm{Z}_{2}^{2}=\frac{\bar{\mathrm{t}}}{4}\bigg(2\xi_1''+\xi_1'^2+\frac{2\xi_1'}{r}\bigg)
+\frac{\bar{\mathrm{t}}'}{2}\bigg(\frac{\xi_1'}{2}+\frac{1}{r}\bigg).
\end{align}

Since we have used the $\mathrm{MGD}$ scheme, both matter sources
must be conserved individually because the exchange of energy is not
permitted in this case. The following two equations confirm the
conservation of these sources as
\begin{align}\label{g22a}
&\frac{dP_r}{dr}+\frac{\xi_1'}{2}\big(\rho+P_r\big)+\frac{2}{r}\big(P_r-P_t\big)
-\frac{\mathrm{qq}'}{4\pi r^4}=0,\\\label{g23a}
&\frac{d\mathrm{Z}_{1}^{1}}{dr}+\frac{\xi_1'}{2}\big(\mathrm{Z}_{1}^{1}-\mathrm{Z}_{0}^{0}\big)
+\frac{2}{r}\big(\mathrm{Z}_{1}^{1}-\mathrm{Z}_{2}^{2}\big)=0.
\end{align}
We observe that the system \eqref{g18}-\eqref{g20} possesses six
unknown quantities ($\rho,P_r,\\P_t,\mathrm{q},\xi_1,\xi_2$),
therefore, we need to choose three of them freely to calculate the
required analytical solution. Further, there are four unknowns
($\bar{\mathrm{t}},\mathrm{Z}_{0}^{0},\mathrm{Z}_{1}^{1},\mathrm{Z}_{2}^{2}$)
in the second set \eqref{g21}-\eqref{g23}. We shall adopt a
constraint on $\mathrm{Z}$-sector to tackle with the second system.
We detect the effective forms of the physical determinants as
\begin{align}\label{g13}
\bar{\rho}&=\rho-\omega\mathrm{Z}_{0}^{0},\\\label{g13a}
\bar{P}_{r}&=P_r+\omega\mathrm{Z}_{1}^{1},\\\label{g13b}
\bar{P}_{t}&=P_t+\omega\mathrm{Z}_{2}^{2},
\end{align}
with the total anisotropy given by
\begin{equation}\label{g14}
\bar{\Pi}=\bar{P}_{t}-\bar{P}_{r}=(P_t-P_r)+\omega(\mathrm{Z}_{2}^{2}-\mathrm{Z}_{1}^{1})=\Pi+\Pi_{\mathrm{Z}}.
\end{equation}
It must be mentioned that the presence of positive or negative
anisotropy can significantly influence the stability of a compact
star. The positive anisotropy (when radial pressure is less than the
tangential component) can increase the stability of a compact star
as it produces outward-directed pressure. This pressure provides a
support against the gravitational attraction and prevents a star
from collapse. On the other hand, the negative anisotropy (when
radial pressure is greater than the tangential component) can lead
to instability of the star as the pressure in the outward direction
is not produced. Thus, such a star remains stable for a short period
of time as compared to that possessing positive anisotropy. Further,
the TOV equation \eqref{g12} can be written as
\begin{align}\label{g14aa}
&\frac{d\bar{P}_r}{dr}+\frac{\xi_1'}{2}\big(\bar{\rho}+\bar{P}_r\big)
+\frac{2}{r}\big(\bar{P}_r-\bar{P}_t\big)-\frac{\mathrm{qq}'}{4\pi
r^4}=0.
\end{align}
which is a combination of different forces. The four terms on the
left side of the above equation represent hydrostatic ($f_{h}$),
gravitational ($f_{g}$), anisotropic ($f_{a}$) and electromagnetic
($f_{e}$) forces, respectively that must be satisfied to maintain
hydrostatic equilibrium of a self-gravitating object. The concise
notation of Eq.\eqref{g14aa} is
\begin{align}\label{g15aa}
&f_h+f_a+f_w=0,
\end{align}
where $f_w=f_g+f_e$.

\section{Isotropization of Anisotropic Compact Fluid Sources}

The anisotropy triggered in a self-gravitating system due to the
original charged fluid source is entirely different from the
anisotropic factor produced by the total configuration (seed and
additional sources). In this section, we provide a brief study that
how the anisotropic interior can be converted into the isotropic
analog. In other words, we find the conditions under which the
considered matter distribution becomes isotropic, i.e.,
$\bar{\Pi}=0$. Following lines show such structural conversion is
being controlled by the decoupling parameter. We observe that the
vanishing decoupling parameter ($\omega=0$) leads to the anisotropic
system, while $\omega=1$ corresponds to the isotropic framework.
Since we are interested in discussing the second case, thus
Eq.\eqref{g14} yields
\begin{equation}\label{g14a}
\Pi_{\mathrm{Z}}=-\Pi \quad \Rightarrow \quad
\mathrm{Z}_{2}^{2}-\mathrm{Z}_{1}^{1}=P_r-P_t.
\end{equation}
The construction of minimally/extended decoupled isotropic interiors
from being anisotropic have been done by taking the above condition
into account \cite{37k,39}.

We now assume a specific metric ansatz to deal with the extra
degrees of freedom in the system \eqref{g18}-\eqref{g20} defined by
\begin{align}\label{g33}
\xi_1(r)&=\ln\bigg\{\mathrm{C}_2^2\bigg(1+\frac{r^2}{\mathrm{C}_1^2}\bigg)\bigg\},\\\label{g34}
\xi_4(r)&=e^{-\xi_2}=\frac{\mathrm{C}_1^2+r^2}{\mathrm{C}_1^2+3r^2},
\end{align}
whose substitution makes the matter triplet as
\begin{align}\label{g35}
\rho&=\frac{6r^4\big(\mathrm{C}_1^2+r^2\big)-\big(\mathrm{C}_1^2+3r^2\big)^2\mathrm{q}^2}{8\pi
r^4\big(\mathrm{C}_1^2+3 r^2\big)^2},\\\label{g35a}
P_r&=\frac{\mathrm{q}^2}{8\pi r^4},\\\label{g35b}
P_t&=\frac{3r^6-\big(\mathrm{C}_1^2+3r^2\big)^2\mathrm{q}^2}{8\pi
r^4 \big(\mathrm{C}_1^2+3r^2\big)^2},
\end{align}
with $\mathrm{C}_1^2$ and $\mathrm{C}_2^2$ being unknown functions
and the junction conditions are used in the following to make them
known. It is important to stress that the metric ansatz \eqref{g33}
and \eqref{g34} correspond only to the tangential pressure in the
uncharged case while the radial component disappears \cite{37k}.
However, they both appear in the current scenario due to the
presence of an electromagnetic field. The same metric potentials
have also been employed in the study of circular-like motion of
different particles in their field of gravitation \cite{42a1}.

Junction conditions are a subject of great discussion of all time
for astrophysicists which assist the study of multiple physical
factors in a self-gravitating interior at the hypersurface, i.e.,
$\Sigma:r=\mathcal{R}$. The smooth matching requires an interior and
exterior metrics representing the corresponding spacetime regions of
the considered geometry. Since we have already defined the interior
metric in Eq.\eqref{g6}, a suitable exterior geometry in this regard
is the Reissner-Nordstr\"{o}m line element (a solution to the
charged vacuum spacetime) given by
\begin{equation}\label{g25}
ds^2=-\left(1-\frac{2\mathcal{M}}{r}+\frac{\mathcal{Q}^2}{r^2}\right)dt^2
+\frac{1}{\left(1-\frac{2\mathcal{M}}{r}+\frac{\mathcal{Q}^2}{r^2}\right)}dr^2+
r^2d\Omega^2,
\end{equation}
with $\mathcal{Q}$ is the total charge and $\mathcal{M}$ symbolizes
the corresponding mass. We now obtain the two constants
\big($\mathrm{C}_1^2,~\mathrm{C}_2^2$\big) by equating the
temporal/radial metrics components of the metrics \eqref{g6} and
\eqref{g25} as
\begin{eqnarray}\label{g37}
\mathrm{C}_1^2&=&\frac{\mathcal{R}^2\big(2\mathcal{R}^2+3\mathcal{Q}^2-6\mathcal{M}\mathcal{R}\big)}
{2\mathcal{M}\mathcal{R}-\mathcal{Q}^2},\\\label{g38}
\mathrm{C}_2^2&=&\frac{2\mathcal{R}^2+3\mathcal{Q}^2-6\mathcal{M}\mathcal{R}}{2\mathcal{R}^2}.
\end{eqnarray}

The radius \big($\mathcal{R}=9.1 \pm 0.4~km$\big) and mass
\big($\mathcal{M}=1.58 \pm 0.06$ times the sun's mass\big) of a
specific compact model $4U~1820-30$ is considered to plot the
resulting solutions in the following sections \cite{42aa}. Further,
we observe that one extra unknown is still present in the system
\eqref{g35}-\eqref{g35b}, thus $\mathrm{q}^2(r)=\xi_5 r^6$ is
assumed with $\xi_5$ as a constant \cite{42aaa}. Joining this with
the restriction \eqref{g14a} and field equations, a differential
equation is obtained as
\begin{align}\nonumber
&r\big(\mathrm{C}_1^2+r^2\big)\big[2r^3\big(\mathrm{C}_1^2+r^2\big)
\big(2\xi_5\mathrm{C}_1^4+12\xi_5\mathrm{C}_1^2r^2+18\xi_5r^4-3\big)-\big(\mathrm{C}_1^2+2r^2\big)\\\label{g39}
&\times\big(\mathrm{C}_1^2+3r^2\big)^2\bar{\mathrm{t}}'(r)\big]+2\big(\mathrm{C}_1^4+2\mathrm{C}_1^2r^2+2r^4\big)
\big(\mathrm{C}_1^2+3r^2\big)^2\bar{\mathrm{t}}(r)=0,
\end{align}
whose exact solution provides $\bar{\mathrm{t}}(r)$ as
\begin{equation}\label{g40}
\bar{\mathrm{t}}(r)=\frac{r^2\big(\mathrm{C}_1^2+r^2\big)}{\mathrm{C}_1^2+2r^2}\bigg\{\mathrm{D}_1
+2\xi_5r^2+\frac{2\xi_5\mathrm{C}_1^2}{3}+\frac{1}{\mathrm{C}_1^2+3r^2}\bigg\},
\end{equation}
with $\mathrm{D}_1$ as an integration constant whose dimension is
$\frac{1}{\ell^2}$. It is well-known that the radial pressure
vanishes at the spherical junction, i.e.,
$\bar{P}_r(\mathcal{R})=0$. Hence, Eq.\eqref{g13a} along with
\eqref{g35a} and \eqref{g40} provides $\mathrm{D}_1$ as
\begin{equation}\label{g40a}
\mathrm{D}_1=-2\xi_5\mathcal{R}^2-\frac{2\xi_5\mathrm{C}_1^2}{3}-\frac{1}{\mathrm{C}_1^2+3\mathcal{R}^2}
-\frac{\xi_5\mathcal{R}^2\big(\mathrm{C}_1^2+2\mathcal{R}^2\big)}{\omega\big(\mathrm{C}_1^2+3\mathcal{R}^2\big)}.
\end{equation}
Equation \eqref{g40} now takes the form
\begin{equation}\label{g40aa}
\bar{\mathrm{t}}(r)=\frac{r^2\big(\mathrm{C}_1^2+r^2\big)}{\mathrm{C}_1^2+2r^2}\bigg\{
2\xi_5\big(r^2-\mathcal{R}^2\big)-\frac{3\big(r^2-\mathcal{R}^2\big)}{\big(\mathrm{C}_1^2+3r^2\big)
\big(\mathrm{C}_1^2+3\mathcal{R}^2\big)}-\frac{\xi_5\mathcal{R}^2\big(\mathrm{C}_1^2+2\mathcal{R}^2\big)}
{\omega\big(\mathrm{C}_1^2+3\mathcal{R}^2\big)}\bigg\},
\end{equation}
and the deformed $g_{rr}$ metric component becomes
\begin{align}\nonumber
e^{\xi_2}=\xi_4^{-1}&=\big[r^2\big(\mathrm{C}_1^2+r^2\big)\big\{-\xi_5\big(\mathrm{C}_1^2+3r^2\big)
\big(\mathrm{C}_1^2\big(\mathcal{R}^2(2\omega+1)-2r^2\omega\big)+2\mathcal{R}^2\\\nonumber
&\times\big(\mathcal{R}^2(3\omega+1)-3r^2\omega\big)\big)-3\omega\big(r^2-\mathcal{R}^2\big)\big\}
+\big(\mathrm{C}_1^2+r^2\big)\big(\mathrm{C}_1^2+2r^2\big)
\\\label{g40aaa} &\times\big(\mathrm{C}_1^2+3\mathcal{R}^2\big)\big]^{-1}\big[\big(\mathrm{C}_1^2+2r^2\big)
\big(\mathrm{C}_1^2+3r^2\big)\big(\mathrm{C}_1^2+3\mathcal{R}^2\big)\big].
\end{align}

Finally, the decoupled solution to the system \eqref{g8}-\eqref{g10}
is expressed by the line element as follows
\begin{equation}\label{g41}
ds^2=-\mathrm{C}_2^2\bigg(1+\frac{r^2}{\mathrm{C}_1^2}\bigg)dt^2+\frac{\mathrm{C}_1^2+3r^2}
{\mathrm{C}_1^2+r^2+\omega\bar{\mathrm{t}}(r)\big(\mathrm{C}_1^2+3r^2\big)}dr^2+
r^2d\Omega^2,
\end{equation}
along with the matter triplet given by
\begin{align}\nonumber
\bar{\rho}&=\frac{\mathrm{C}_1^2\big(6-6\xi_5r^4\big)-\mathrm{C}_1^4\xi_5r^2-9\xi_5r^6+6r^2}{8\pi\big(\mathrm{C}_1^2+3r^2\big)^2}
-\big[8 \pi  r^2 \big(\mathrm{C}_1^2+r^2\big)^2
\big\{\mathrm{C}_1^4\big(\xi _5 r^2\\\nonumber &\times \big(2 r^2
\omega -\mathcal{R}^2 (2 \omega +1)\big)+1\big)+\mathrm{C}_1^2
\big(2 r^2+\xi _5 r^2 \big(6 r^4 \omega -3 r^2 \mathcal{R}^2-2
\mathcal{R}^4\\\nonumber &\times (3 \omega +1)\big)+3
\mathcal{R}^2\big)+3 r^2 \big(2 \xi _5 r^2 \mathcal{R}^2 \big(3 r^2
\omega -\mathcal{R}^2 (3 \omega +1)\big)-\omega
r^2+\mathcal{R}^2\\\nonumber &\times(\omega
+2)\big)\big\}^2\big]^{-1}\big[\omega \big(\mathrm{C}_1^2+3
\mathcal{R}^2\big) \big\{\mathrm{C}_1^{10}\big(\xi _5 r^2
\big(\mathcal{R}^2 (2 \omega +1)-6 r^2 \omega \big)+1\big)
\\\nonumber &+\mathrm{C}_1^8 \big(3 \big(4
r^2+\mathcal{R}^2\big)+\xi _5 r^2 \big(-50 r^4 \omega +r^2
\mathcal{R}^2 (7-4 \omega )+2 \mathcal{R}^4 (3 \omega
+1)\big)\big)\\\nonumber &+\mathrm{C}_1^6 r^2 \big(r^2 (9 \omega
+39)+\xi _5 r^2 \big(r^2 \mathcal{R}^2 (17-116 \omega )+14
\mathcal{R}^4 (3 \omega +1)-150 r^4 \omega\big)\\\nonumber &-3
\mathcal{R}^2 (\omega -12)\big)+\mathrm{C}_1^4 r^4 \big(r^2 (30
\omega +44)+\xi _5 r^2 \big(3 r^2 \mathcal{R}^2 (7-136 \omega )-198
r^4 \omega \\\nonumber &+34 \mathcal{R}^4 (3 \omega +1)\big)+3
\mathcal{R}^2 (2 \omega +39)\big)+3 \mathrm{C}_1^2 r^6 \big(r^2 (9
\omega +4)+2 \xi _5 r^2 \big(-18 r^4 \omega \\\nonumber &+3 r^2
\mathcal{R}^2 (1-31 \omega )+7 \mathcal{R}^4 (3 \omega
+1)\big)+\mathcal{R}^2 (13 \omega +44)\big)+18 r^8 \big(2 \xi _5 r^2
\mathcal{R}^2 \big(\mathcal{R}^2\\\label{g46} &\times (3 \omega
+1)-9 r^2 \omega \big)+r^2 \omega +\mathcal{R}^2 (\omega
+2)\big)\big\}\big] ,\\\nonumber
\bar{P}_{r}&=\frac{1}{8\pi}\bigg[\xi_5r^2+\omega\bigg(\frac{\mathrm{C}_1^2+3r^2}{\mathrm{C}_1^2+2r^2}\bigg)
\bigg\{2\xi_5\big(r^2-\mathcal{R}^2\big)-\frac{3\big(r^2-\mathcal{R}^2\big)}{\big(\mathrm{C}_1^2+3r^2\big)
\big(\mathrm{C}_1^2+3\mathcal{R}^2\big)}\\\label{g47}
&-\frac{\xi_5\mathcal{R}^2\big(\mathrm{C}_1^2+2\mathcal{R}^2\big)}
{\omega\big(\mathrm{C}_1^2+3\mathcal{R}^2\big)}\bigg\}\bigg],\\\nonumber
\bar{P}_{t}&=\frac{r^2\big(3-\mathrm{C}_1^4\xi_5-6\mathrm{C}_1^2\xi_5
r^2-9\xi_5r^4\big)}{8\pi\big(\mathrm{C}_1^2+3r^2\big)^2}+\frac{\omega
r^2\big(2\mathrm{C}_1^2+r^2\big)}{8\pi\big(\mathrm{C}_1^2+r^2\big)\big(\mathrm{C}_1^2+2r^2\big)}\\\nonumber
&\times\bigg\{2\xi_5\big(r^2-\mathcal{R}^2\big)-\frac{3\big(r^2-\mathcal{R}^2\big)}{\big(\mathrm{C}_1^2+3r^2\big)
\big(\mathrm{C}_1^2+3\mathcal{R}^2\big)}-\frac{\xi_5\mathcal{R}^2\big(\mathrm{C}_1^2+2\mathcal{R}^2\big)}
{\omega\big(\mathrm{C}_1^2+3\mathcal{R}^2\big)}\bigg\}-\frac{\omega}{8\pi}\\\nonumber
&\times\big[\big(\mathrm{C}_1^2+r^2\big)^3 \big(\xi_5 r^2
\big(\mathrm{C}_1^2+3 r^2\big) \big(\mathrm{C}_1^2 \big(2 r^2 \omega
-\mathcal{R}^2 (2 \omega +1)\big)+2 \mathcal{R}^2 \big(3 r^2
\omega\\\nonumber &-\mathcal{R}^2 (3 \omega
+1)\big)\big)+\mathrm{C}_1^2 \big(\mathrm{C}_1^2+2 r^2+3
\mathcal{R}^2\big)+3 r^2 \big(\mathcal{R}^2 (\omega +2)-r^2 \omega
\big)\big)^2\big]^{-1}\\\nonumber &\times\big[\big(\mathrm{C}_1^2+2
r^2\big) \big(\mathrm{C}_1^2+3 \mathcal{R}^2\big) \big\{-2
\mathrm{C}_1^4 r^2 \big(4 r^2 (3 \omega +1)-3 \mathcal{R}^2 (\omega
-4)\big)-\mathrm{C}_1^6 \\\nonumber &\times \big(r^2 (6 \omega +8)-3
\mathcal{R}^2 (\omega -2)\big)-3 \mathrm{C}_1^2 r^4 \big(10 r^2
\omega +\mathcal{R}^2 (\omega +8)\big)+\xi _5
\big(\mathrm{C}_1^2\\\nonumber &+3 r^2\big)^2 \big(\mathrm{C}_1^6
\big(-\big(\mathcal{R}^2 (2 \omega +1)-4 r^2 \omega \big)\big)+2
\mathrm{C}_1^4 \big(5 r^4 \omega +r^2 \mathcal{R}^2 (4 \omega
-1)\mathcal{R}^4\\\nonumber &-\times (3 \omega +1)\big)+2
\mathrm{C}_1^2 r^2 \big(4 r^4 \omega +r^2 \mathcal{R}^2 (13 \omega
-1)-2 \mathcal{R}^4 (3 \omega +1)\big)+4 r^4 \mathcal{R}^2
\\\label{g48} &\times\big(6r^2 \omega-\mathcal{R}^2 (3 \omega +1)\big)\big)-2 \mathrm{C}_1^8-18
r^8 \omega \big\}\big].
\end{align}
The anisotropy in the interior of the above developed model is
\begin{eqnarray}\label{g49}
\bar{\Pi}&=&\frac{r^2\big(3-2\xi_5\mathrm{C}_1^4-12\xi_5\mathrm{C}_1^2r^2-18\xi_5r^4\big)}{8\pi\big(\mathrm{C}_1^2+3r^2\big)^2}
\big(1-\omega\big).
\end{eqnarray}
It becomes clear from the above equation that the anisotropy fades
away for $\omega=1$, hence, the total matter distribution turns into
the isotropic interior for this particular value. We can now say
that Eqs.\eqref{g46}-\eqref{g49} are the analytical solution of the
Einstein-Maxwell field equations for $\omega \in [0,1]$. In other
words, the variation in this parameter produces the isotropic
configuration from being anisotropic and vice versa.

\section{Complexity Analysis and Compact Fluid Sources}

Herrera \cite{37g} defined complexity for the first time in such a
way that could be suitable for all scientific fields. According to
this definition, a uniform/ homogenous system is always
complexity-free, implying that the density inhomogeneity and
anisotropy in the pressure make the structure complex. Multiple
structure scalars, in this context, were obtained corresponding to a
static spherical interior through the orthogonal decomposition of
the curvature tensor. Both the above-mentioned factors were found in
one of the scalars, i.e., $\mathrm{Y}_{TF}$ and thus entitled the
complexity factor. This work was also extended for a non-static
scenario where some evolutionary patterns have been discussed
\cite{37h}. For the charged scenario, the factor $\mathrm{Y}_{TF}$
becomes
\begin{equation}\label{g51}
\mathrm{Y}_{TF}(r)=8\pi\Pi+\frac{4\mathrm{q}^2}{r^4}-\frac{4\pi}{r^3}\int_0^r\bar{y}^3\rho'(\bar{y})d\bar{y}
-\frac{3}{r^3}\int_0^r\frac{\mathrm{qq}'}{\bar{y}}d\bar{y}.
\end{equation}
Since the current setup \eqref{g8}-\eqref{g10} involves two matter
sources, thus the corresponding extension of the complexity factor
is given by
\begin{eqnarray}\nonumber
\bar{\mathrm{Y}}_{TF}(r)&=&8\pi\bar{\Pi}+\frac{4\mathrm{q}^2}{r^4}-\frac{4\pi}{r^3}\int_0^r\bar{y}^3\bar{\rho}'(\bar{y})d\bar{y}
-\frac{3}{r^3}\int_0^r\frac{\mathrm{qq}'}{\bar{y}}d\bar{y}\\\nonumber
&=&8\pi\Pi+\frac{4\mathrm{q}^2}{r^4}-\frac{4\pi}{r^3}\int_0^r\bar{y}^3\rho'(\bar{y})d\bar{y}
-\frac{3}{r^3}\int_0^r\frac{\mathrm{qq}'}{\bar{y}}d\bar{y}\\\label{g54}
&+&8\pi\omega\Pi_{\mathrm{Z}}+\frac{4\pi\omega}{r^3}\int_0^r\bar{y}^3\mathrm{Z}{_0^0}'(\bar{y})d\bar{y},
\end{eqnarray}
which can also be written as
\begin{eqnarray}\label{g55}
\bar{\mathrm{Y}}_{TF}=\mathrm{Y}_{TF}+\mathrm{Y}_{TF}^{\mathrm{Z}}.
\end{eqnarray}
Here, $\mathrm{Y}_{TF}$ and $\mathrm{Y}_{TF}^{\mathrm{Z}}$ are the
complexity factors for the sources \eqref{g18}-\eqref{g20} and
\eqref{g21}-\eqref{g23}, respectively. Since we develop the model
\eqref{g46}-\eqref{g49} by taking $\bar{\Pi}=0$ into account,
therefore, Eq.\eqref{g54} leads to
\begin{eqnarray}\label{g56}
\bar{\mathrm{Y}}_{TF}&=&\frac{4\mathrm{q}^2}{r^4}-\frac{4\pi}{r^3}\int_0^r\bar{y}^3\bar{\rho}'(\bar{y})d\bar{y}
-\frac{3}{r^3}\int_0^r\frac{\mathrm{qq}'}{\bar{y}}d\bar{y}.
\end{eqnarray}
After substituting the derivative of the effective energy density
\eqref{g46} in the above equation, we get
\begin{align}\nonumber
\bar{\mathrm{Y}}_{TF}&=\frac{11\xi_5r^2}{5}+\frac{r^2}{5\big(\mathrm{C}_1^4+5\mathrm{C}_1^2r^2+6r^4\big)^2
\big(\mathrm{C}_1^2+3\mathcal{R}^2\big)}\big[\mathrm{C}_1^{10} (10
\xi_5  \omega +\xi_5 )+2 \xi_5\\\nonumber &\times  \mathrm{C}_1^8
\big(5 (8 \omega+1)r^2+4(5\omega+1)\mathcal{R}^2\big)+\mathrm{C}_1^6
\big(\xi_5 (230 \omega +37) r^4+60 \xi_5r^2 \mathcal{R}^2
\\\nonumber & (5 \omega
+1)+5 \big(2 \xi_5 (3 \omega +1) \mathcal{R}^4+6-3 \omega
\big)\big)+6 \mathrm{C}_1^4 \big(10 r^6 (5 \xi_5 \omega +\xi_5 )+26
\xi_5 \\\nonumber &\times(5 \omega +1) r^4 \mathcal{R}^2+5 r^2
\big(2 \xi_5 (3 \omega +1) \mathcal{R}^4+4-\omega \big)+5 (3-2
\omega ) \mathcal{R}^2\big)+3 \mathrm{C}_1^2 \\\nonumber &\times r^2
\big(12 r^6 (5 \xi_5 \omega +\xi_5 )+60 \xi_5 (5 \omega +1) r^4
\mathcal{R}^2-60 (\omega -2) \mathcal{R}^2+5 r^2 \big(\omega +6
\\\label{g56a} &\times \xi_5 (3 \omega +1) \mathcal{R}^4+8\big)\big)+18 r^4
\mathcal{R}^2 \big(6 r^4 (5 \xi_5 \omega +\xi_5 )-5 (\omega
-4)\big)\big].
\end{align}

\subsection{Complexity-free additional Matter Source}

Here we consider the additional fluid source to be the
complexity-free (i.e., $\mathrm{Y}_{TF}^{\mathrm{Z}}=0$) as an extra
constraint so that the system \eqref{g21}-\eqref{g23} can be solved
uniquely. After engaging this with Eq.\eqref{g55}, we observe that
$\bar{\mathrm{Y}}_{TF}=\mathrm{Y}_{TF}$ or, equivalently
\begin{eqnarray}\label{g56b}
\Pi_{\mathrm{Z}}=-\frac{1}{2r^3}\int_0^r\bar{y}^3\mathrm{Z}{_0^0}'(\bar{y})d\bar{y}.
\end{eqnarray}
The integral on the right side of Eq.\eqref{g56b} can be manipulated
by using \eqref{g21} as
\begin{eqnarray}\label{g56c}
\int_0^r\bar{y}^3\mathrm{Z}{_0^0}'(\bar{y})d\bar{y}=
r^2\bar{\mathrm{t}}'(r)-2r\bar{\mathrm{t}}(r),
\end{eqnarray}
whose substitution along with Eqs.\eqref{g22} and \eqref{g23} in
\eqref{g56b} results in the first-order differential equation as
follows
\begin{eqnarray}\label{g56d}
&&\bar{\mathrm{t}}'(r)\bigg(\frac{\xi_1'}{4}+\frac{1}{r}\bigg)+\bar{\mathrm{t}}(r)\bigg(\frac{\xi_1''}{2}-\frac{2}{r^2}
+\frac{\xi_1'^2}{4}-\frac{\xi_1'}{2r}\bigg)=0.
\end{eqnarray}
The above equation contains the metric coefficient, thus the
solution for $\bar{\mathrm{t}}(r)$ is possible only if we consider a
particular metric ansatz. In order to continue our study, we take
Tolman IV metric components given in the following
\begin{align}\label{g57}
\xi_1(r)&=\ln\bigg\{\mathrm{C}_2^2\bigg(1+\frac{r^2}{\mathrm{C}_1^2}\bigg)\bigg\},\\\label{g58}
\xi_4(r)&=e^{-\xi_2(r)}=\frac{\big(\mathrm{C}_1^2+r^2\big)
\big(\mathrm{C}_3^2-r^2\big)}{\mathrm{C}_3^2\big(\mathrm{C}_1^2+2r^2\big)},
\end{align}
characterizing the isotropic interior through the following density
and pressure as
\begin{align}\label{g59}
\rho&=\frac{\mathrm{C}_1^4\big(3-\xi_5
r^2\mathrm{C}_3^2\big)+\mathrm{C}_1^2\big(7r^2-4\xi_5
r^4\mathrm{C}_3^2+3\mathrm{C}_3^2\big)-4\xi_5
r^6\mathrm{C}_3^2+6r^4+2r^2\mathrm{C}_3^2}{8\pi\mathrm{C}_3^2\big(\mathrm{C}_1^2+2r^2\big)^2},\\\label{g60}
P&=\frac{\xi_5\mathrm{C}_1^2r^2\mathrm{C}_3^2-\mathrm{C}_1^2+2\xi_5
r^4\mathrm{C}_3^2-3r^2+\mathrm{C}_3^2}{8\pi\mathrm{C}_1^2\mathrm{C}_3^2+16\pi
r^2\mathrm{C}_3^2}.
\end{align}

The values of $\mathrm{C}_1^2$ and $\mathrm{C}_2^2$ are already
obtained in Eqs.\eqref{g37} and \eqref{g38}, however, the newly
introduced constant $\mathrm{C}_3^2$ becomes
\begin{align}\label{g60a}
\mathrm{C}_3^2&=\frac{2\mathcal{R}^3\big(\mathcal{Q}^2-2\mathcal{M}\mathcal{R}+\mathcal{R}^2\big)}
{2\mathcal{M}\mathcal{Q}-4\mathcal{M}^2\mathcal{R}+\mathcal{Q}\mathcal{R}+2\mathcal{M}\mathcal{R}^2}.
\end{align}
Inserting Eq.\eqref{g57} in \eqref{g56d} and simplifying, we obtain
\begin{align}\label{g60aa}
r\big(2\mathrm{C}_1^4+5\mathrm{C}_1^2r^2+3r^4\big)\bar{\mathrm{t}}'(r)
-2\big(2\mathrm{C}_1^4+4\mathrm{C}_1^2r^2+3r^4\big)\bar{\mathrm{t}}(r)=0,
\end{align}
whose solution is
\begin{equation}\label{g60b}
\bar{\mathrm{t}}(r)=\frac{\mathrm{D}_2r^2\big(\mathrm{C}_1^2+r^2\big)}{2\mathrm{C}_1^2+3r^2},
\end{equation}
along with a constant $\mathrm{D}_2$ having dimension of
$\frac{1}{\ell^2}$. We again use
$\bar{P}_r(\mathcal{R})=P_r(\mathcal{R})+\omega\mathrm{Z}_1^1(\mathcal{R})=0$
to calculate $\mathrm{D}_2$ and simplification gives
\begin{equation}\label{g60ba}
\mathrm{D}_2=\frac{\big(\mathrm{C}_1^2-\mathrm{C}_3^2+3\mathcal{R}^2-\xi_5\mathcal{R}^2\mathrm{C}_1^2\mathrm{C}_3^2
-2\xi_5\mathcal{R}^4\mathrm{C}_3^2\big)\big(2\mathrm{C}_1^2+3\mathcal{R}^2\big)}{\omega
\mathrm{C}_3^2\big(\mathrm{C}_1^2+2\mathcal{R}^2\big)\big(\mathrm{C}_1^2+3\mathcal{R}^2\big)}.
\end{equation}
Consequently, the deformed expression of $g_{rr}$ component becomes
\begin{align}\label{g60c}
e^{\xi_2}&=\xi_4^{-1}=\frac{\mathrm{C}_3^2\big(\mathrm{C}_1^2+2r^2\big)\big(2\mathrm{C}_1^2+3r^2\big)}{\big(\mathrm{C}_1^2+r^2\big)
\big\{\mathrm{C}_3^2\big(r^2\omega\mathrm{D}_2\big(\mathrm{C}_1^2+2r^2\big)+2\mathrm{C}_1^2+3r^2\big)-r^2
\big(2\mathrm{C}_1^2+3r^2\big)\big\}}.
\end{align}
Since the constraint \eqref{g56b} is based only on the additional
matter source that does not have the influence of charge, the
deformation function \eqref{g60b} is identical with the uncharged
scenario \cite{37k}. Therefore, we do not need to physically
interpret the corresponding model as this has already been done.
Moreover, the factor $\bar{\mathrm{Y}}_{TF}$ \big(expressed in
Eq.\eqref{g54}\big) now becomes
\begin{align}\label{g60d}
\bar{\mathrm{Y}}_{TF}=\mathrm{Y}_{TF}=\frac{r^2\big(\mathrm{C}_1^2+2\mathrm{C}_3^2)}
{\mathrm{C}_3^2\big(\mathrm{C}_1^2+2r^2\big)^2}+\frac{2\xi_5r^2}{5}.
\end{align}
The above factor does not contain the decoupling parameter, thus we
adopt the deformed $g_{rr}$ metric component \eqref{g60c} to obtain
the constant $\mathrm{C}_3^2$ so that the variation of
$\bar{\mathrm{Y}}_{TF}$ with respect to $\omega$ can be shown. We
get this constant as
\begin{align}\nonumber
\mathrm{C}_3^2&=2\mathcal{R}^4\big(6\mathcal{M}\mathcal{R}-3\mathcal{Q}^2-4\mathcal{R}^2\big)\big[2\mathrm{D}_2\mathcal{R}^4
\omega\big(2\mathcal{M}\mathcal{R}-\mathcal{Q}^2-2\mathcal{R}^2\big)\\\label{g60da}
&+4\mathcal{M}\mathcal{R}^2\big(3\mathcal{M}-2\mathcal{R}\big)+\mathcal{Q}^2\big(3\mathcal{Q}^2
-12\mathcal{M}\mathcal{R}+4\mathcal{R}^2\big)\big]^{-1}.
\end{align}

\subsection{Complexity-free Total Matter Source}

In this subsection, we try another constraint on the newly added
source to make the Einstein-Maxwell field equations solvable so that
a physically relevant compact interior can be modeled. We assume
that the seed and additional sources may possess complexity
individually, i.e., $\mathrm{Y}_{TF}\neq0$ and
$\mathrm{Y}_{TF}^{\mathrm{Z}}\neq0$, however, the system is no more
complex once both sources merged, or equivalently,
$\bar{\mathrm{Y}}_{TF}=0$. This assumption makes the scalar
\eqref{g55} in terms of Tolman IV ansatz as
\begin{align}\nonumber
&r \big(\mathrm{C}_1^2+r^2\big) \big[5 \mathrm{C}_3^2
\big(\mathrm{C}_1^2+2 r^2\big)^2 \big(2 \mathrm{C}_1^2+3 r^2\big)
\bar{\mathrm{t}}'(r)+2 r^3 \big(\mathrm{C}_1^2+r^2\big)\\\nonumber &
\big\{2 \mathrm{C}_3^2 \mathrm{C}_1^4 \xi _5+2 \mathrm{C}_3^2 \big(4
\xi _5 r^4+5\big)+\mathrm{C}_1^2 \big(8 \mathrm{C}_3^2 \xi _5
r^2+5\big)\big\}\big]\\\label{60e} &-10 \mathrm{C}_3^2
\big(\mathrm{C}_1^2+2 r^2\big)^2 \big(4 \mathrm{C}_1^2 r^2+2
\mathrm{C}_1^4+3 r^4\big) \bar{\mathrm{t}}(r)=0,
\end{align}
providing the function $\bar{\mathrm{t}}(r)$ as follows
\begin{align}\label{60f}
\bar{\mathrm{t}}(r)&=\frac{r^2\big(\mathrm{C}_1^2+r^2\big)}{5\big(2\mathrm{C}_1^2+3r^2\big)}
\bigg\{5\bigg(\frac{1}{\mathrm{C}_1^2+2r^2}+\mathrm{D}_3\bigg)-\mathrm{C}_1^2\xi_5+\frac{5\mathrm{C}_1^2}{\mathrm{C}_3^2
\big(2\mathrm{C}_1^2+4r^2\big)}-2\xi_5r^2\bigg\},
\end{align}
where an arbitrary constant $\mathrm{D}_3$ has a dimension
$\frac{1}{\ell^2}$. Using Eq.\eqref{g13a} at the spherical junction,
we get the value $\mathrm{D}_3$ as
\begin{align}\nonumber
\mathrm{D}_3&=\frac{1}{5\mathrm{C}_3^2\big(\mathrm{C}_1^2+2\mathcal{R}^2\big)}
\bigg[\xi_5\mathrm{C}_3^2\big(\mathrm{C}_1^2+2\mathcal{R}^2\big)^2-5\mathrm{C}_3^2-\frac{5\mathrm{C}_1^2}{2}\\\label{60faa}
&+\frac{5\big\{\mathrm{C}_1^2-\mathrm{C}_3^2+3\mathcal{R}^2-\xi_5\mathcal{R}^2\mathrm{C}_3^2\big(\mathrm{C}_1^2
+2\mathcal{R}^2\big)\big\}\big(2\mathrm{C}_1^2+3\mathcal{R}^2\big)}{\omega\big(\mathrm{C}_1^2+3\mathcal{R}^2\big)}\bigg].
\end{align}
Putting this back into Eq.\eqref{60f}, we obtain
\begin{align}\nonumber
\bar{\mathrm{t}}(r)&=\frac{r^2\big(\mathrm{C}_1^2+r^2\big)}{10
\mathrm{C}_3^2 \big(2 \mathrm{C}_1^2+3 r^2\big)
\big(\mathrm{C}_1^2+2 \mathcal{R}^2\big)}\bigg[\big(\mathrm{C}_1^2+2
\mathcal{R}^2\big) \bigg\{\frac{5 \mathrm{C}_1^2}{\mathrm{C}_1^2+2
r^2}-2 \mathrm{C}_3^2 \xi _5\\\nonumber &\times
\big(\mathrm{C}_1^2+2 r^2\big)\bigg\}+\frac{10 \mathrm{C}_3^2
\big(\mathrm{C}_1^2+2 \mathcal{R}^2\big)}{\mathrm{C}_1^2+2 r^2}+2
\mathrm{C}_3^2 \xi _5 \big(\mathrm{C}_1^2+2 \mathcal{R}^2\big)^2-10
\mathrm{C}_3^2\\\label{60faaa} &-5 \mathrm{C}_1^2+\frac{10 \big(2
\mathrm{C}_1^2+3 \mathcal{R}^2\big) \big(\mathrm{C}_3^2 \xi _5
\mathcal{R}^2 \big(-\big(\mathrm{C}_1^2+2
\mathcal{R}^2\big)\big)+\mathrm{C}_1^2-\mathrm{C}_3^2+3
\mathcal{R}^2\big)}{\omega \big(\mathrm{C}_1^2+3
\mathcal{R}^2\big)}\bigg].
\end{align}

The corresponding modified form of $g_{rr}$ metric potential can be
expressed through the transformation \eqref{g17} as
\begin{align}\nonumber
e^{\xi_2}=\xi_4^{-1}&=\frac{10 \mathrm{C}_3^2 \big(\mathrm{C}_1^2+2
r^2\big) \big(2 \mathrm{C}_1^2+3 r^2\big) \big(5 \mathrm{C}_1^2
\mathcal{R}^2+\mathrm{C}_1^4+6
\mathcal{R}^4\big)}{r^2+\mathrm{C}_1^2}\\\nonumber &\times\big[10
\big(2 \mathrm{C}_1^2+3 r^2\big) \big(\mathrm{C}_3^2-r^2\big) \big(5
\mathrm{C}_1^2 \mathcal{R}^2+\mathrm{C}_1^4+6 \mathcal{R}^4\big)-2
r^2 \big\{5 \mathrm{C}_3^2 \big(\mathrm{C}_1^2 \\\nonumber
&\times\big(2 r^2(\omega +2)+\mathcal{R}^2 (3-2 \omega )\big)+2
\mathrm{C}_1^4+6 \mathcal{R}^2 \big(r^2 (\omega +1)-\mathcal{R}^2
\omega \big)\big)\\\nonumber &+5 \big(\mathrm{C}_1^2+3
\mathcal{R}^2\big) \big(\mathrm{C}_1^2 \big(r^2 (\omega
-4)-\mathcal{R}^2 (\omega +3)\big)-2 \mathrm{C}_1^4-6 r^2
\mathcal{R}^2\big)\\\nonumber &+\mathrm{C}_3^2 \xi _5
\big(\mathrm{C}_1^2+2 r^2\big) \big(\mathrm{C}_1^2+2
\mathcal{R}^2\big) \big(2 \mathrm{C}_1^2 \big(r^2 \omega
+\mathcal{R}^2 (5-\omega)\big)+6 r^2 \mathcal{R}^2
\omega\\\label{g60fa} &+\mathcal{R}^4 (15-6 \omega
)\big)\big\}\big]^{-1}.
\end{align}
Hence, the corresponding matter determinants take the final form as
\begin{align}\nonumber
\bar{\rho}&=\frac{1}{40 \pi  \mathrm{C}_3^2 \big(7 \mathrm{C}_1^2
r^2+2 \mathrm{C}_1^4+6 r^4\big)^2 \big(5 \mathrm{C}_1^2
\mathcal{R}^2+\mathrm{C}_1^4+6 \mathcal{R}^4\big)}\bigg[4
\mathrm{C}_3^2 \mathrm{C}_1^{12} \xi _5\big\{(\omega -1)\\\nonumber
&\times5 r^2 +3 \mathcal{R}^2 (5-\omega )\big\}+36 \mathrm{C}_3^2
r^6 \mathcal{R}^2 \big\{2 \xi _5 r^2 \mathcal{R}^2 \big(5 r^2 (2
\omega -3)+3 \mathcal{R}^2 (5-2 \omega )\big)\\\nonumber &+5 \big(3
r^2 (\omega +1)+\mathcal{R}^2 (3-\omega )\big)\big\}+2
\mathrm{C}_1^{10} \big\{\mathrm{C}_3^2 \big(\xi _5 \big(7 r^4 (9
\omega -10)+(13 \omega +135)\\\nonumber &\times r^2 \mathcal{R}^2
+15 \mathcal{R}^4 (7-2 \omega )\big)+60\big)+5 (\omega -1) \big(5
r^2-3 \mathcal{R}^2\big)\big\}+6 \mathrm{C}_1^2 r^4
\big\{\mathrm{C}_3^2 \big(2 \xi _5 \\\nonumber &\times r^2
\mathcal{R}^2 \big(25 r^4 (2 \omega -3)+r^2 \mathcal{R}^2 (122
\omega -105)+44 \mathcal{R}^4 (5-2 \omega )\big)+5 \big( (\omega
+2)\\\nonumber &\times6 r^4+r^2 \mathcal{R}^2 (39 \omega +59)+3
\mathcal{R}^4 (17-5 \omega )\big)\big)+15 r^2 \mathcal{R}^2
(\omega-1) \big(3r^2-\mathcal{R}^2\big)\big\}\\\nonumber
&+\mathrm{C}_1^8 \big\{\mathrm{C}_3^2 \big(10 \big(r^2 (10 \omega
+59)+3 \mathcal{R}^2 (13-2 \omega )\big)+\xi _5 \big(r^6 (294 \omega
-365)+10 \mathcal{R}^2\\\nonumber &\times r^4(46 \omega +15)+25 r^2
\mathcal{R}^4 (47-10 \omega )+36 \mathcal{R}^6 (5-2 \omega
)\big)\big)+5 (\omega -1) \big(35 r^4\\\nonumber &+13 r^2
\mathcal{R}^2-18 \mathcal{R}^4\big)\big\}+\mathrm{C}_1^6
\big\{\mathrm{C}_3^2 \big(5 \big(36 \mathcal{R}^4 (2-\omega )+r^2
\mathcal{R}^2 (26 \omega +331)+r^4 \\\nonumber &\times (70 \omega
+221)\big)+\xi _5 r^2 \big(4 r^6 (76 \omega -105)+r^4 \mathcal{R}^2
(1294 \omega -945)+r^2 \mathcal{R}^4 (2135\\\nonumber &-94 \omega
)+222 \mathcal{R}^6 (5-2 \omega )\big)\big)+5 r^2 (\omega -1)
\big(41 r^4+90 r^2 \mathcal{R}^2-51
\mathcal{R}^4\big)\big\}+\mathrm{C}_1^4 r^2\\\nonumber &\times
\big\{2 \mathrm{C}_3^2 \big(5 \big(r^4 (41 \omega +97)+15 r^2
\mathcal{R}^2 (6 \omega +17)+3 \mathcal{R}^4 (44-17 \omega
)\big)+\xi _5 r^2 \big(30 r^6\\\nonumber &\times (2 \omega -3)+2 r^4
\mathcal{R}^2 (362 \omega -435)+r^2 \mathcal{R}^4 (442 \omega
+445)+255 \mathcal{R}^6 (5-2 \omega )\big)\big)\\\label{60g} &+45
r^2 (\omega -1) \big(2 r^4+13 r^2 \mathcal{R}^2-5
\mathcal{R}^4\big)\big\}\bigg],\\\nonumber
\bar{P}_{r}&=\frac{1}{80\pi \mathrm{C}_3^2 \big(2 \mathrm{C}_1^2+3
r^2\big) \big(\mathrm{C}_1^2+2 \mathcal{R}^2\big)} \bigg[\bigg(10
\mathrm{C}_3^2 \xi _5 r^2-\frac{10
\big(\mathrm{C}_1^2-\mathrm{C}_3^2+3 r^2\big)}{\mathrm{C}_1^2+2
r^2}\bigg)\\\nonumber &\times \big(2 \mathrm{C}_1^2+3 r^2\big)
\big(\mathrm{C}_1^2+2 \mathcal{R}^2\big) +\omega
\big(\mathrm{C}_1^2+3r^2\big) \bigg\{\frac{10 \mathrm{C}_3^2
\big(\mathrm{C}_1^2+2 \mathcal{R}^2\big)}{\mathrm{C}_1^2+2 r^2}-5
\mathrm{C}_1^2+2 \xi_5\\\nonumber &\times \mathrm{C}_3^2
\big(\mathrm{C}_1^2+2 \mathcal{R}^2\big)^2-10
\mathrm{C}_3^2+\big(\mathrm{C}_1^2+2 \mathcal{R}^2\big)
\bigg(\frac{5 \mathrm{C}_1^2}{\mathrm{C}_1^2+2 r^2}-2 \mathrm{C}_3^2
\xi _5 \big(\mathrm{C}_1^2+2 r^2\big)\bigg)\\\label{60h} &+\frac{10
\big(2 \mathrm{C}_1^2+3 \mathcal{R}^2\big) \big(\mathrm{C}_3^2 \xi
_5 \mathcal{R}^2
\big(-\big(\mathrm{C}_1^2+2\mathcal{R}^2\big)\big)+\mathrm{C}_1^2-\mathrm{C}_3^2+3
\mathcal{R}^2\big)}{\omega \big(\mathrm{C}_1^2+3
\mathcal{R}^2\big)}\bigg\}\bigg],
\\\nonumber
\bar{P}_{t}&=\frac{-1}{40 \pi  \mathrm{C}_3^2 \big(\mathrm{C}_1^2+2
r^2\big) \big(2 \mathrm{C}_1^2+3 r^2\big)^2 \big(5 \mathrm{C}_1^2
\mathcal{R}^2+\mathrm{C}_1^4+6 \mathcal{R}^4\big)}\bigg[4
\mathrm{C}_3^2 \mathrm{C}_1^{10}\big\{5 \xi_5 r^2\\\nonumber &+\xi
_5 \big(2 r^2 \omega -\mathcal{R}^2 (\omega -5)\big)\big\}+18
\mathrm{C}_3^2 r^4 \mathcal{R}^2 \{2 \xi _5 r^2 \mathcal{R}^2\big(10
r^2 \omega +\mathcal{R}^2 (15-6 \omega )\big)
\\\nonumber &+5 \big(6 \xi_5
r^4 \mathcal{R}^2+3 r^2 (\omega +1)-\mathcal{R}^2 (\omega
+3)\big)\big\}+2 \mathrm{C}_1^8 \big\{\mathrm{C}_3^2 \big(50 \xi_5
r^2 \big(r^2+\mathcal{R}^2\big)\\\nonumber &+\xi _5 \big(25 r^4
\omega +2 r^2 \mathcal{R}^2 (3 \omega +35)+5 \mathcal{R}^4 (7-2
\omega )\big)\big)+5 (\omega -1) \big(2
r^2-\mathcal{R}^2\big)\big\}\\\nonumber &+3 \mathrm{C}_1^2 r^2
\big\{2 \mathrm{C}_3^2 \big(\xi _5 r^2 \mathcal{R}^2 \big(50 r^4
\omega +r^2 \mathcal{R}^2 (68 \omega +105)+29 \mathcal{R}^4 (5-2
\omega )\big)+5\\\nonumber &\times \big(15 \xi_5  r^6
\mathcal{R}^2+3 r^4 \big(11 \xi_5 \mathcal{R}^4+\omega +2\big)+r^2
\mathcal{R}^2 (12 \omega +7)-6 \mathcal{R}^4 (\omega
+2)\big)\big)\\\nonumber &+15 r^2 \mathcal{R}^2 (\omega -1) \big(3
r^2-\mathcal{R}^2\big)\big\} +\mathrm{C}_1^6 \big\{\mathrm{C}_3^2
\big(5 \big(33 \xi_5  r^6+100 \xi_5  r^4 \mathcal{R}^2+8 r^2
\big(2\\\nonumber &+3 \xi_5 \mathcal{R}^4+\omega \big)-2
\mathcal{R}^2 (2 \omega +7)\big)+2 \xi _5 \big(49 r^6 \omega +r^4
\mathcal{R}^2 (96 \omega +145)+r^2 \mathcal{R}^4 \\\nonumber
&\times(245-46 \omega )+6 \mathcal{R}^6 (5-2 \omega )\big)\big)+5
(\omega -1) \big(13 r^4+6 r^2 \mathcal{R}^2-6
\mathcal{R}^4\big)\big\}+\mathrm{C}_1^4
\\\nonumber &\times\big\{\mathrm{C}_3^2 \big(\xi _5 r^2\big(60 r^6
\omega +2 r^4 \mathcal{R}^2 (227 \omega +90)+5 r^2 \mathcal{R}^4 (2
\omega +203)+ (5-2 \omega )84 \\\nonumber
&\times\mathcal{R}^6\big)+5 \big(18 \xi_5 r^8+165 \xi_5 r^6
\mathcal{R}^2+r^4 \big(120 \xi_5 \mathcal{R}^4+26 \omega +49\big)+6
r^2 \mathcal{R}^2 \\\label{60i} &\times (2 \omega -3)-12
\mathcal{R}^4 (\omega +2)\big)\big)+45 r^2 (\omega -1) \big(r^4+4
r^2 \mathcal{R}^2-2 \mathcal{R}^4\big)\big\}\bigg].
\end{align}
Further, Eqs.\eqref{60h} and \eqref{60i} can be used to obtain the
corresponding pressure anisotropy.

\section{Graphical Description of the Newly Obtained Models}

The mass function corresponding to the charged spherical geometry is
defined in Eq.\eqref{g12b}. We put the effective energy densities
\eqref{g46} and \eqref{60g} to observe how this factor behaves
against $r$ for our developed models. This function also helps to
define the compactness and redshift of a self-gravitating system.
The former parameter \big(symbolizes by $\tau(r)$\big) reveals the
tightness of particles that how they arrange in a specific style. In
other words, we can define it as the ratio of mass and radius of a
compact body, thus an increasing function of $r$ outwards. Several
researchers have been tried to find its lower and upper bounds in a
physically acceptable interior and the maximum value was found to be
$\frac{4}{9}$ at the boundary \cite{42a}. Another parameter is known
as the redshift which expresses the change in the wavelength of
electromagnetic radiations ejecting from a massive object influenced
by the strong gravitational field that nearby system produces. The
mathematical notation is given by
\begin{equation}
z(r)=\frac{1-\sqrt{1-2\tau(r)}}{\sqrt{1-2\tau(r)}},
\end{equation}
whose upper bound has been suggested in the literature as $2$ and
$5.211$ for the case of perfect \cite{42a} and anisotropic
\cite{42b} matter sources, respectively.

A subject of scientific debate for astrophysicists is to check the
existence of an ordinary fluid in a compact interior. Multiple
constraints, in this regard, gained much significance because they
are used to check the physical viability of the models. These
constraints are referred as the energy conditions, and their
satisfaction (dissatisfaction of any of them) verifies the presence
of usual (exotic) fluid. They are, in fact, linear combination of
different physical determinants characterizing the interior of a
compact model. The influence of an electromagnetic field leads these
bounds to the following
\begin{eqnarray}\nonumber
&&\bar{\rho}+\frac{\mathrm{q}^2}{8\pi r^4} \geq 0, \quad
\bar{\rho}+\bar{P}_{r} \geq 0,\\\nonumber
&&\bar{\rho}+\bar{P}_{t}+\frac{\mathrm{q}^2}{4\pi r^4} \geq 0, \quad
\bar{\rho}-\bar{P}_{r}+\frac{\mathrm{q}^2}{4\pi r^4} \geq
0,\\\label{g50} &&\bar{\rho}-\bar{P}_{t} \geq 0, \quad
\bar{\rho}+\bar{P}_{r}+2\bar{P}_{t}+\frac{\mathrm{q}^2}{4\pi r^4}
\geq 0.
\end{eqnarray}

The most important phenomenon in the study of compact stars is that
how long these structures remain stable during evolutionary changes.
Therefore, we employ sound speed criteria to check the stable
regions of our proposed models. The sound speed is defined as the
variation in pressure against the variation in the density, i.e.,
$v_{s}^{2}=\frac{dP}{d\rho}$. Since the current scenario contains
effective variables as well as anisotropic fluid, the sound speed is
classified in radial
$\big(v_{sr}^{2}=\frac{d\bar{P}_{r}}{d\bar{\rho}}\big)$ and
tangential $\big(v_{st}^{2}=\frac{d\bar{P}_{t}}{d\bar{\rho}}\big)$
directions. Their acceptable ranges were reported as $0 <
v_{sr}^{2},~ v_{st}^{2} < 1$ to get a stable system \cite{42bb}.
Herrera then studied the occurrence of cracking in compact stars and
combined the above both factors in a single framework, and suggested
that a stable interior must fulfill $0 < |v_{st}^{2}-v_{sr}^{2}| <
1$ \cite{42ba}. Another phenomenon in this regard is the adiabatic
index that plays a key role in determining the nature of pressure.
According to the research \cite{44}, if the adiabatic index is
greater than $\frac{4}{3}$, a perturbation that compresses the star
will lead to an increase in pressure. This increment in pressure
resists the compression and ultimately contributes to the star's
stability. On the other hand, if the adiabatic index is less than
the above-mentioned limit, compression will cause a decrease in
pressure, leading to further compression and overall instability of
the star occurs. Mathematically, it is defined as
\begin{equation}\label{g62}
\bar{\Gamma}=\frac{\bar{\rho}+\bar{P}_{r}}{\bar{P}_{r}}
\left(\frac{d\bar{P}_{r}}{d\bar{\rho}}\right)=\frac{\bar{\rho}+\bar{P}_{r}}{\bar{P}_{r}}
\left(v_{sr}^{2}\right).
\end{equation}

We now discuss physical interpretation of the obtained models by
plotting the corresponding deformation functions, extended $g_{rr}$
metric potentials, physical parameters and several other factors. We
adopt four different values of the decoupling parameter such as
$\omega=0.25,0.5,0.75$ and $1$. Further, the effect of charge on
developed models is also checked by taking $\mathcal{Q}=0.1,0.8$ and
$\xi_5=-0.001$ into account. The first model corresponds to
$\bar{\Pi}=0$ is investigated in the following. Figure \textbf{1}
(left) exhibits the deformation function \eqref{g40aa} that
disappears at $r=0$, increasing initially and then decreasing
towards the spherical boundary. The right plot indicates that the
corresponding radial coefficient \eqref{g40aaa} shows a non-singular
increasing trend, and it takes the same values at the core as well
as hypersurface for all parametric values. The physical determinants
must be maximum in the center of a star and decrease towards its
boundary, as indicated by Figure \textbf{2}. The upper left graph
shows that the energy density decreases with a rise in the
decoupling parameter and charge near the center, and vice versa near
the boundary. The tangential pressure, on the other hand, behaves
opposite to the energy density (lower left). Further, the spherical
junction possesses only the tangential pressure whereas the radial
component becomes null at that point. As far as the anisotropy is
concerned, it becomes zero at the core (as radial and tangential
pressures are equal) and increases outwards (lower right). It must
be mentioned that the above-described behavior of anisotropy is just
for $\omega=0.25,0.5$ and $0.75$. However, this factor vanishes
throughout for $\omega=1$, leading to the isotropic interior.
\begin{figure}\center
\epsfig{file=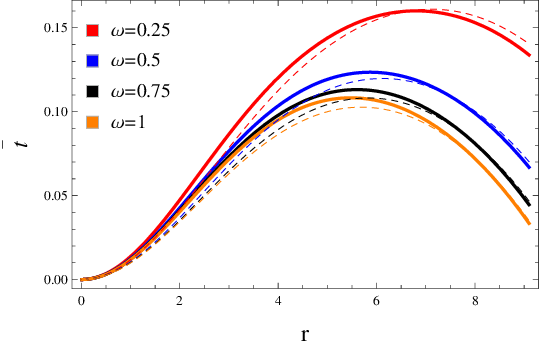,width=0.4\linewidth}\epsfig{file=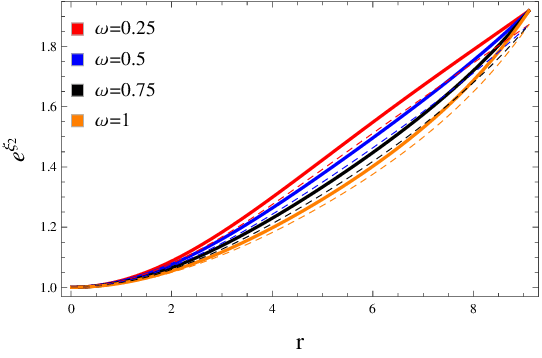,width=0.4\linewidth}
\caption{Deformation function \eqref{g40aa} and deformed $g_{rr}$
metric potential \eqref{g40aaa} corresponding to $\mathcal{Q}=0.1$
(solid) and $0.8$ (dashed) for the solution generated by
$\bar{\Pi}=0$.}
\end{figure}
\begin{figure}\center
\epsfig{file=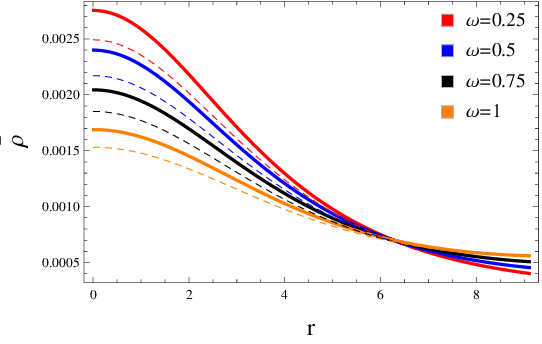,width=0.4\linewidth}\epsfig{file=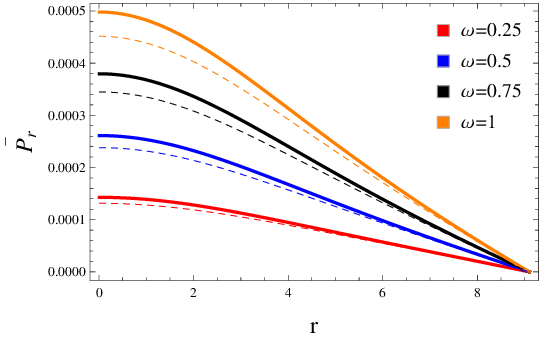,width=0.4\linewidth}
\epsfig{file=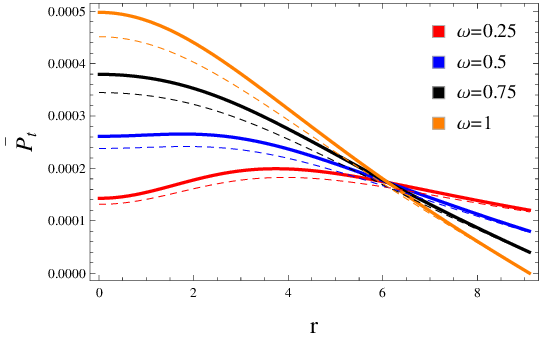,width=0.4\linewidth}\epsfig{file=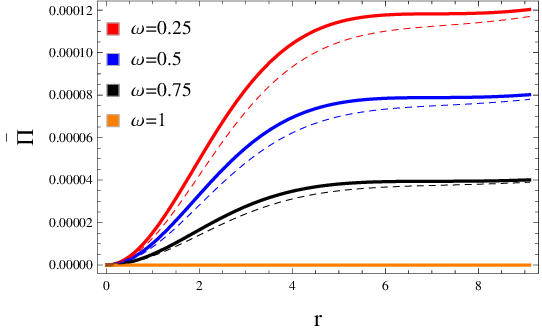,width=0.4\linewidth}
\caption{Physical variables corresponding to $\mathcal{Q}=0.1$
(solid) and $0.8$ (dashed) for the solution generated by
$\bar{\Pi}=0$.}
\end{figure}

The mass function is plotted in Figure \textbf{3} (upper left) that
becomes zero at $r=0$ and possesses an increasing trend outwards. It
is observed that the anisotropic interior is dense as compared to
the isotropic analog for both values of the electric charge. The
upper right and lower graphs are in agreement with the required
criteria of compactness and redshift. Tables \textbf{1} and
\textbf{2} provide the numerical values of these quantities.
Further, we notice that the density and both pressures are
positively definite everywhere, thus only the dominant energy bounds
like $\bar{\rho}-\bar{P}_{r}+\frac{\mathrm{q}^2}{4\pi r^4} \geq 0$
and $\bar{\rho}-\bar{P}_{t} \geq 0$ are plotted in Figure
\textbf{4}, making sure the existence of viable resulting model.
\begin{table}[H]
\scriptsize \centering \caption{Values of physical parameters for a
compact star $4U ~1820-30$ with $\mathcal{Q}=0.1$ corresponding to
the solution generated by $\bar{\Pi}=0$.} \label{Table1}
\vspace{+0.1in} \setlength{\tabcolsep}{0.95em}
\begin{tabular}{cccccc}
% after \\: \hline or \cline{col1-col2} \cline{col3-col4} ...
\hline\hline $\omega$ & $\rho_c~(gm/cm^3)$ & $\rho_s~(gm/cm^3)$ &
$P_{c}~(dyne/cm^2)$ & $\tau_s$ & $z_s$
\\\hline
$0.25$ & 3.6844$\times$10$^{15}$ & 5.4369$\times$10$^{14}$ &
1.7267$\times$10$^{35}$ & 0.247 & 0.406
\\\hline
$0.5$ & 3.1988$\times$10$^{15}$ & 5.9748$\times$10$^{14}$ &
3.1095$\times$10$^{35}$ & 0.247 & 0.406
\\\hline
$0.75$ & 2.7332$\times$10$^{15}$ & 6.6932$\times$10$^{14}$ &
4.5608$\times$10$^{35}$ & 0.247 & 0.406
\\\hline
$1$ & 2.2663$\times$10$^{15}$ & 7.5909$\times$10$^{14}$ & 5.9772$\times$10$^{35}$ & 0.247 & 0.406 \\
\hline\hline
\end{tabular}
\end{table}
\begin{table}[H]
\scriptsize \centering \caption{Values of physical parameters for a
compact star $4U ~1820-30$ with $\mathcal{Q}=0.8$ corresponding to
the solution generated by $\bar{\Pi}=0$.} \label{Table2}
\vspace{+0.1in} \setlength{\tabcolsep}{0.95em}
\begin{tabular}{cccccc}
% after \\: \hline or \cline{col1-col2} \cline{col3-col4} ...
\hline\hline $\omega$ & $\rho_c~(gm/cm^3)$ & $\rho_s~(gm/cm^3)$ &
$P_{c}~(dyne/cm^2)$ & $\tau_s$ & $z_s$
\\\hline
$0.25$ & 3.3245$\times$10$^{15}$ & 5.4289$\times$10$^{14}$ &
1.5535$\times$10$^{35}$ & 0.238 & 0.384
\\\hline
$0.5$ & 2.8937$\times$10$^{15}$ & 6.1755$\times$10$^{14}$ &
2.8329$\times$10$^{35}$ & 0.238 & 0.384
\\\hline
$0.75$ & 2.4817$\times$10$^{15}$ & 6.7762$\times$10$^{14}$ &
4.1808$\times$10$^{35}$ & 0.238 & 0.384
\\\hline
$1$ & 2.0509$\times$10$^{15}$ & 7.4932$\times$10$^{14}$ & 5.4241$\times$10$^{35}$ & 0.238 & 0.384 \\
\hline\hline
\end{tabular}
\end{table}
\begin{figure}\center
\epsfig{file=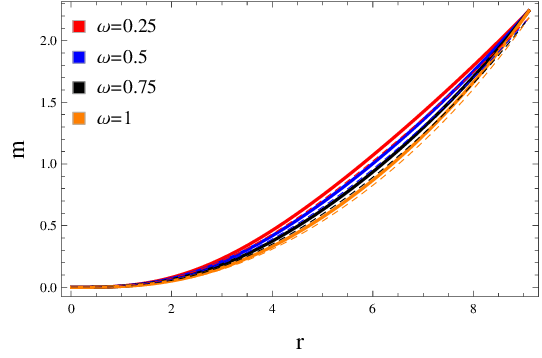,width=0.4\linewidth}\epsfig{file=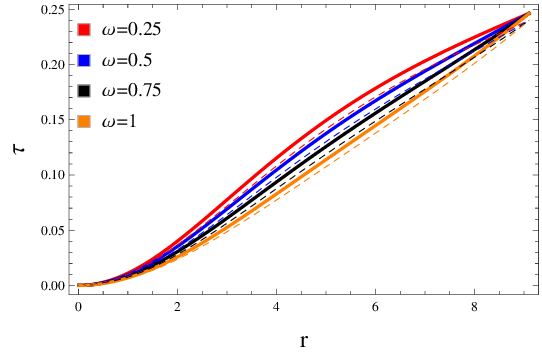,width=0.4\linewidth}
\epsfig{file=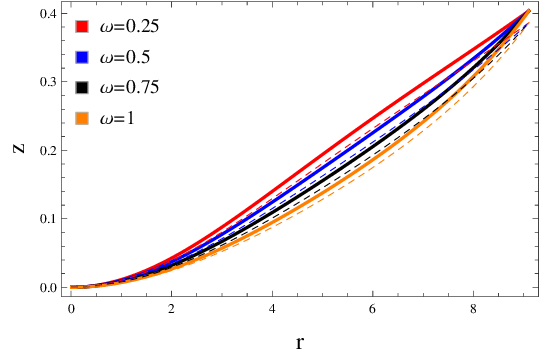,width=0.4\linewidth} \caption{Different
factors corresponding to $\mathcal{Q}=0.1$ (solid) and $0.8$
(dashed) for the solution generated by $\bar{\Pi}=0$.}
\end{figure}
\begin{figure}\center
\epsfig{file=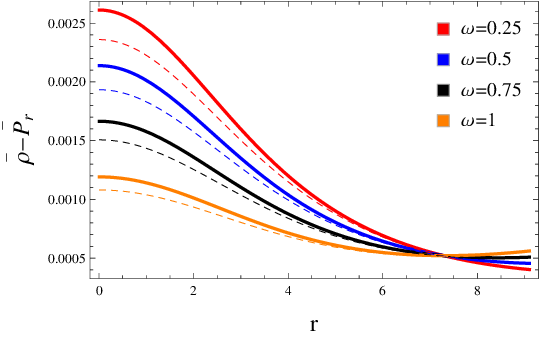,width=0.4\linewidth}\epsfig{file=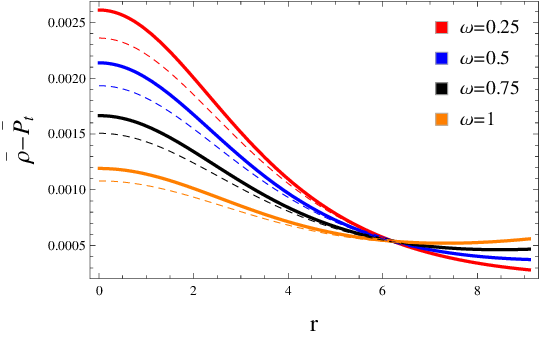,width=0.4\linewidth}
\caption{Dominant energy bounds corresponding to $\mathcal{Q}=0.1$
(solid) and $0.8$ (dashed) for the solution generated by
$\bar{\Pi}=0$.}
\end{figure}

Different forces in TOV equation \eqref{g15aa} are plotted in Figure
\textbf{5} and we find that the resulting solution is in the
hydrostatic equilibrium. Figure \textbf{6} determines the stability
analysis by means of sound speeds and adiabatic index. The radial
and tangential sound speed components alongside the lower right plot
provide that our model is stable for all parametric values except
$\omega=1$. This implies that the isotropic analog is no more
physically relevant in contrast with \cite{37m}. The scalars
\eqref{g56a} and \eqref{g60d} describing the complexity of compact
sources are plotted in Figure \textbf{7} which decrease and increase
with the increment in $\omega$, respectively. However, the first
scalar increases throughout with the radial coordinate while the
other factor initially increases and then decreases towards the
spherical junction. Further, the reduction in the complexity is
observed in the presence of charge, making it interesting to be
studied.
\begin{figure}\center
\epsfig{file=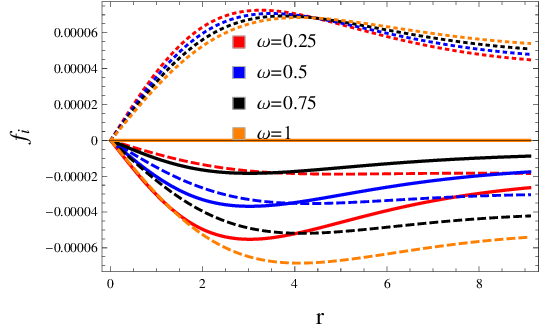,width=0.4\linewidth}\epsfig{file=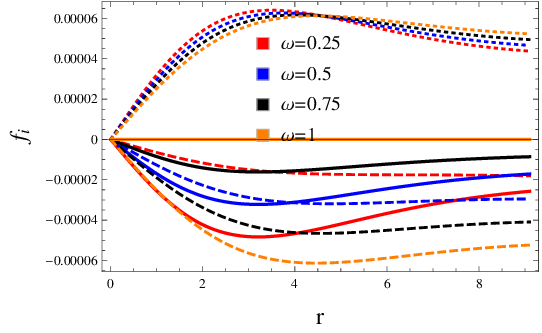,width=0.4\linewidth}
\caption{Different forces including $f_a$ (solid), $f_h$ (dashed)
and $f_w$ (dotted) corresponding to $\mathcal{Q}=0.1$ (left) and
$0.8$ (right) for the solution generated by $\bar{\Pi}=0$.}
\end{figure}
\begin{figure}\center
\epsfig{file=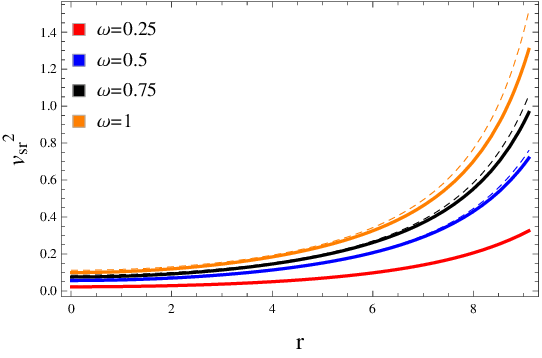,width=0.4\linewidth}\epsfig{file=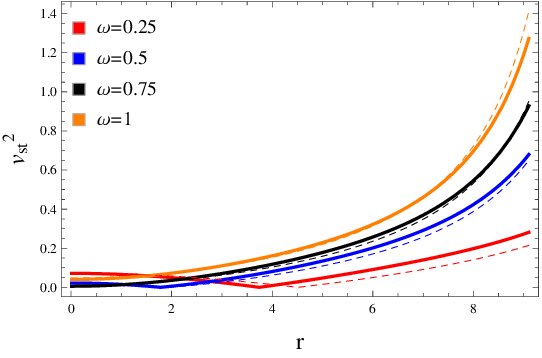,width=0.4\linewidth}
\epsfig{file=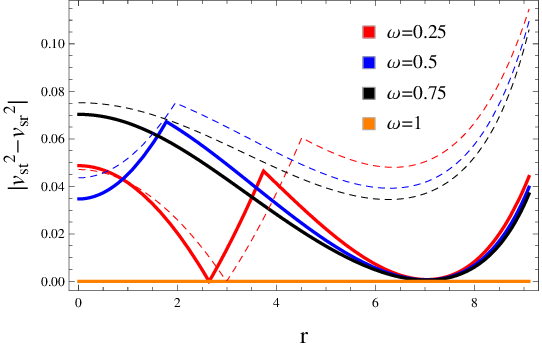,width=0.4\linewidth}\epsfig{file=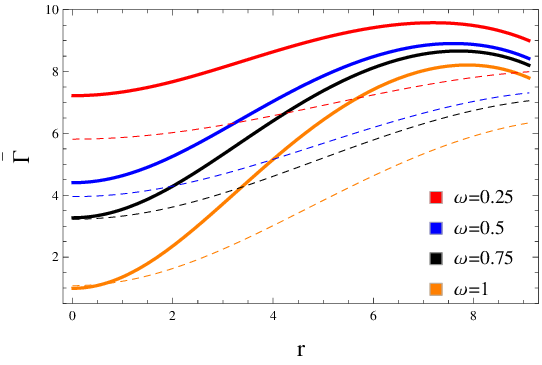,width=0.4\linewidth}
\caption{Radial/tangential speeds of sound, $|v_{st}^2-v_{sr}^2|$
and adiabatic index corresponding to $\mathcal{Q}=0.1$ (solid) and
$0.8$ (dashed) for the solution generated by $\bar{\Pi}=0$.}
\end{figure}
\begin{figure}\center
\epsfig{file=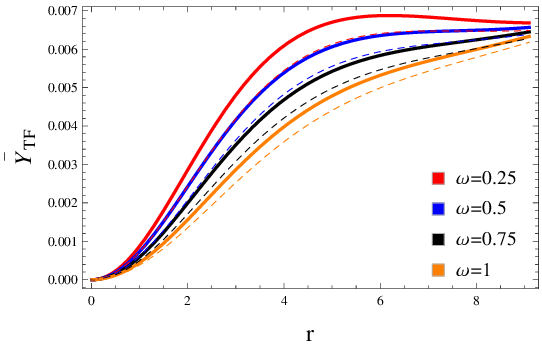,width=0.4\linewidth}\epsfig{file=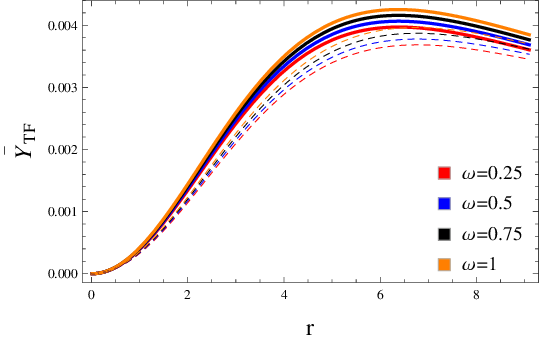,width=0.4\linewidth}
\caption{Complexity factors \eqref{g56a} and \eqref{g60d}
corresponding to $\mathcal{Q}=0.1$ (solid) and $0.8$ (dashed).}
\end{figure}

The deformation functions and their corresponding radial metric
components for the constraints $\mathrm{Y}_{TF}^{\mathrm{Z}}=0$ and
$\bar{\mathrm{Y}}_{TF}=0$ are pictured in Figures \textbf{8} and
\textbf{9}, respectively, showing an acceptable (increasing and free
from singularity) trend. The solution for the later constraint is
now physically interpreted in the following for the same parametric
choices. Figure \textbf{10} demonstrates the profile of the matter
triplet \big(presented in Eqs.\eqref{60g}-\eqref{60i}\big) along
with the corresponding pressure anisotropy. The density and pressure
components exhibit the same behavior as we have already observed
corresponding to the first model. However, the anisotropy initially
becomes zero at the core, decreases outwards against $r$ and then
again possesses increasing behavior. Further, the less value of
charge makes the structure more anisotropic (lower right). Figure
\textbf{11} (upper left) points out that there is a slight
difference between the spherical mass function for different values
of $\omega$. Moreover, other criteria for redshift and compactness
are also fulfilled. The numerical values of these physical factors
are presented in Tables \textbf{3} and \textbf{4}. The acceptable
behavior of energy conditions ensures a viable resulting model that
can be seen in Figure \textbf{12}. The hydrostatic equilibrium
condition for this model is verified in Figure \textbf{13} for all
parametric values. The upper left and lower right plots of Figure
\textbf{14} reveal that the developed solution is stable for every
parametric choice except $\omega=1$.
\begin{table}[H]
\scriptsize \centering \caption{Values of physical parameters for a compact star $4U
~1820-30$ with $\mathcal{Q}=0.1$ corresponding to the solution generated by
$\mathrm{Y}_{TF}^{\mathrm{Z}}=0$.} \label{Table3} \vspace{+0.1in}
\setlength{\tabcolsep}{0.95em}
\begin{tabular}{cccccc}
% after \\: \hline or \cline{col1-col2} \cline{col3-col4} ...
\hline\hline $\omega$ & $\rho_c~(gm/cm^3)$ & $\rho_s~(gm/cm^3)$ & $P_{c}~(dyne/cm^2)$
& $\tau_s$ & $z_s$
\\\hline
$0.25$ & 2.3426$\times$10$^{15}$ &
6.6624$\times$10$^{14}$ & 5.7295$\times$10$^{35}$ &
0.248 & 0.404
\\\hline
$0.5$ & 2.1472$\times$10$^{15}$ &
6.8417$\times$10$^{14}$ & 6.2935$\times$10$^{35}$ &
0.248 & 0.404
\\\hline
$0.75$ & 1.9425$\times$10$^{15}$ &
7.2096$\times$10$^{14}$ & 6.9007$\times$10$^{35}$ &
0.248 & 0.404
\\\hline
$1$ & 1.7512$\times$10$^{15}$ & 7.5401$\times$10$^{14}$ & 7.5079$\times$10$^{35}$ & 0.248 & 0.404 \\
\hline\hline
\end{tabular}
\end{table}
\begin{table}[H]
\scriptsize \centering \caption{Values of physical parameters for a compact star $4U
~1820-30$ with $\mathcal{Q}=0.8$ corresponding to the solution generated by
$\mathrm{Y}_{TF}^{\mathrm{Z}}=0$.} \label{Table4} \vspace{+0.1in}
\setlength{\tabcolsep}{0.95em}
\begin{tabular}{cccccc}
% after \\: \hline or \cline{col1-col2} \cline{col3-col4} ...
\hline\hline $\omega$ & $\rho_c~(gm/cm^3)$ & $\rho_s~(gm/cm^3)$ & $P_{c}~(dyne/cm^2)$
& $\tau_s$ & $z_s$
\\\hline
$0.25$ & 2.1606$\times$10$^{15}$ &
6.5193$\times$10$^{14}$ & 5.0791$\times$10$^{35}$ &
0.239 & 0.386
\\\hline
$0.5$ & 1.9773$\times$10$^{15}$ &
6.8257$\times$10$^{14}$ & 5.5997$\times$10$^{35}$ &
0.239 & 0.386
\\\hline
$0.75$ & 1.8088$\times$10$^{15}$ &
7.1013$\times$10$^{14}$ & 6.1636$\times$10$^{35}$ &
0.239 & 0.386
\\\hline
$1$ & 1.6322$\times$10$^{15}$ & 7.4183$\times$10$^{14}$ & 6.7275$\times$10$^{35}$ & 0.239 & 0.386 \\
\hline\hline
\end{tabular}
\end{table}
\begin{figure}\center
\epsfig{file=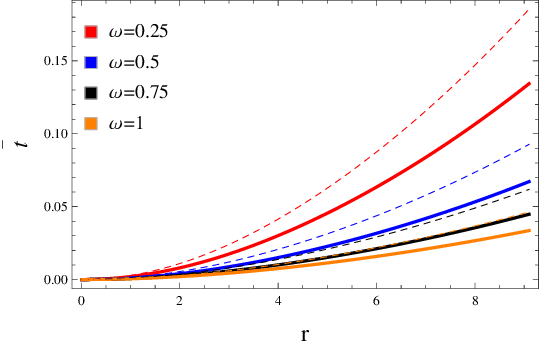,width=0.4\linewidth}\epsfig{file=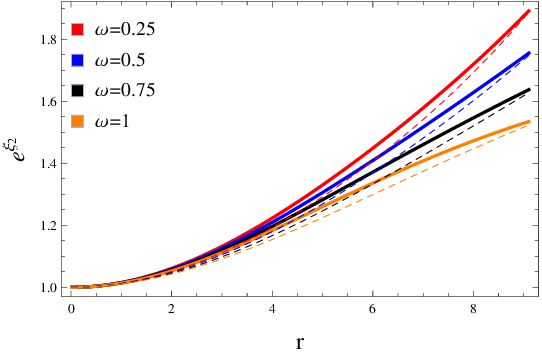,width=0.4\linewidth}
\caption{Deformation function \eqref{g60b} and deformed $g_{rr}$
metric potential \eqref{g60c} corresponding to $\mathcal{Q}=0.1$
(solid) and $0.8$ (dashed) for the solution generated by
$\mathrm{Y}_{TF}^{\mathrm{Z}}=0$.}
\end{figure}
\begin{figure}\center
\epsfig{file=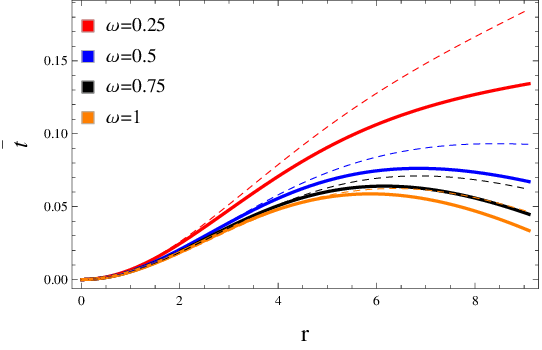,width=0.4\linewidth}\epsfig{file=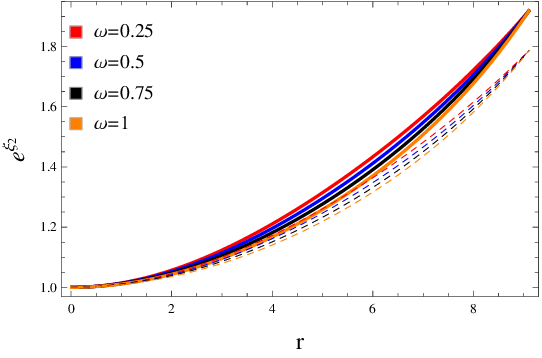,width=0.4\linewidth}
\caption{Deformation function \eqref{60faaa} and deformed $g_{rr}$
metric potential \eqref{g60fa} corresponding to $\mathcal{Q}=0.1$
(solid) and $0.8$ (dashed) for the solution generated by
$\bar{\mathrm{Y}}_{TF}=0$.}
\end{figure}
\begin{figure}\center
\epsfig{file=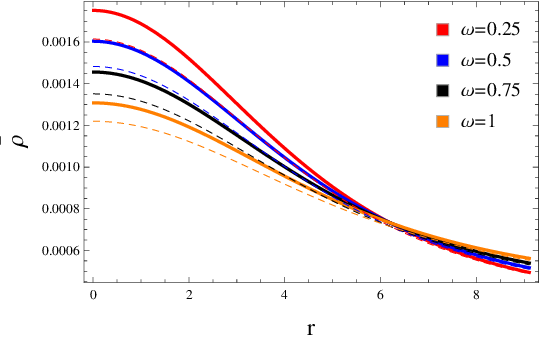,width=0.4\linewidth}\epsfig{file=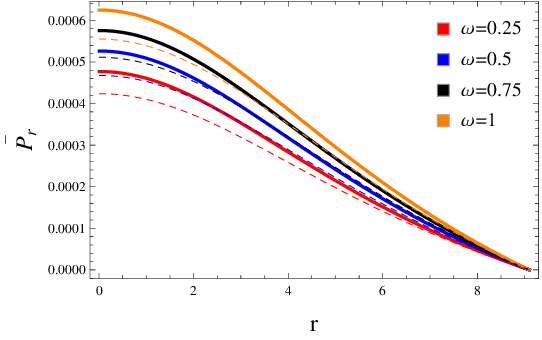,width=0.4\linewidth}
\epsfig{file=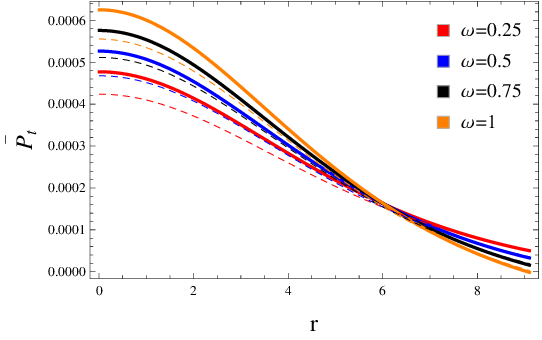,width=0.4\linewidth}\epsfig{file=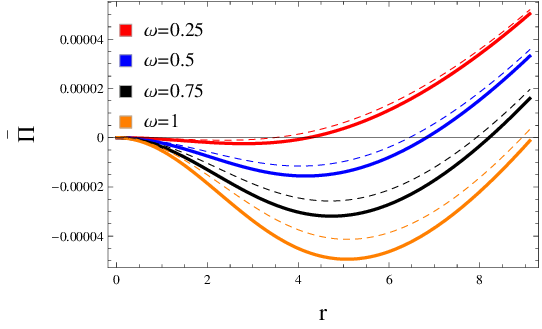,width=0.4\linewidth}
\caption{Physical variables corresponding to $\mathcal{Q}=0.1$
(solid) and $0.8$ (dashed) for the solution generated by
$\bar{\mathrm{Y}}_{TF}=0$.}
\end{figure}
\begin{figure}\center
\epsfig{file=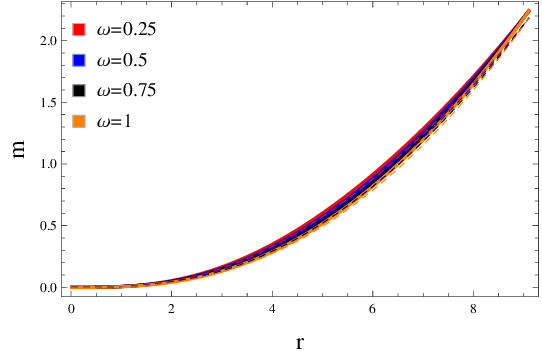,width=0.4\linewidth}\epsfig{file=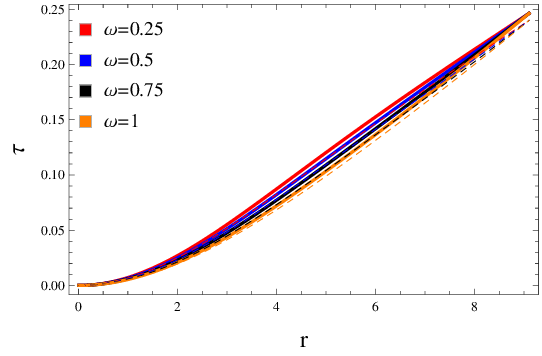,width=0.4\linewidth}
\epsfig{file=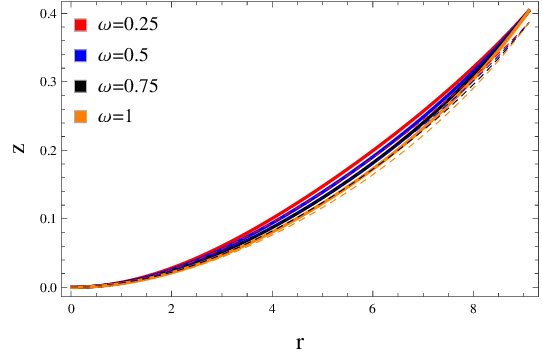,width=0.4\linewidth} \caption{Different
factors corresponding to $\mathcal{Q}=0.1$ (solid) and $0.8$
(dashed) for the solution generated by $\bar{\mathrm{Y}}_{TF}=0$.}
\end{figure}
\begin{figure}\center
\epsfig{file=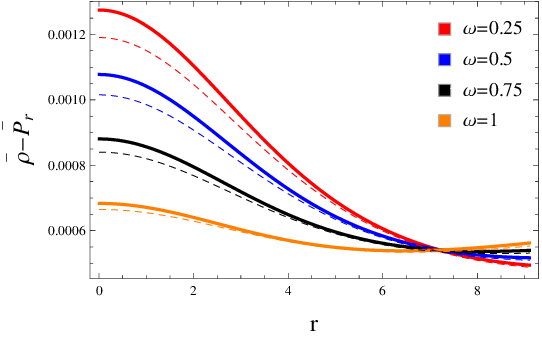,width=0.4\linewidth}\epsfig{file=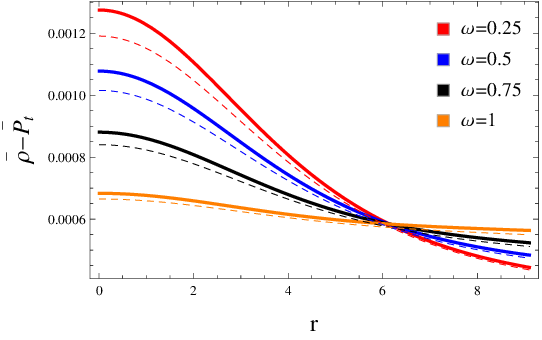,width=0.4\linewidth}
\caption{Dominant energy bounds corresponding to $\mathcal{Q}=0.1$
(solid) and $0.8$ (dashed) for the solution generated by
$\bar{\mathrm{Y}}_{TF}=0$.}
\end{figure}
\begin{figure}\center
\epsfig{file=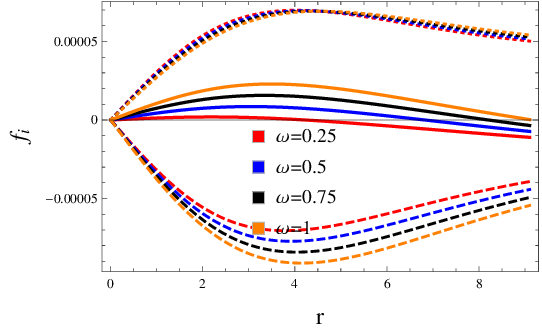,width=0.4\linewidth}\epsfig{file=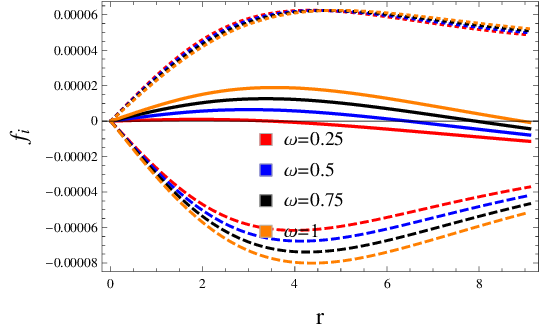,width=0.4\linewidth}
\caption{Different forces including $f_a$ (solid), $f_h$ (dashed)
and $f_w$ (dotted) corresponding to $\mathcal{Q}=0.1$ (left) and
$0.8$ (right) for the solution generated by
$\bar{\mathrm{Y}}_{TF}=0$.}
\end{figure}
\begin{figure}\center
\epsfig{file=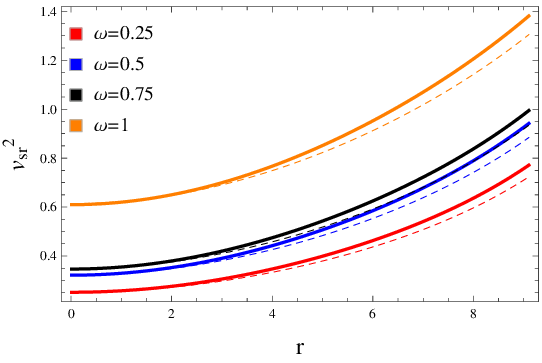,width=0.4\linewidth}\epsfig{file=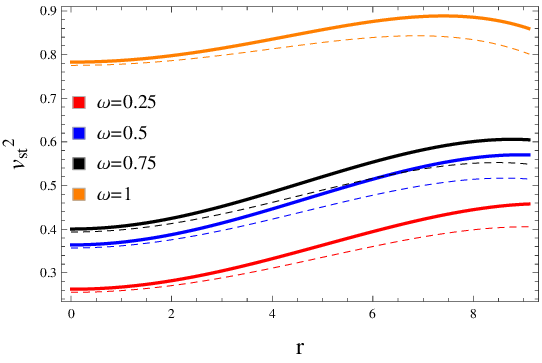,width=0.4\linewidth}
\epsfig{file=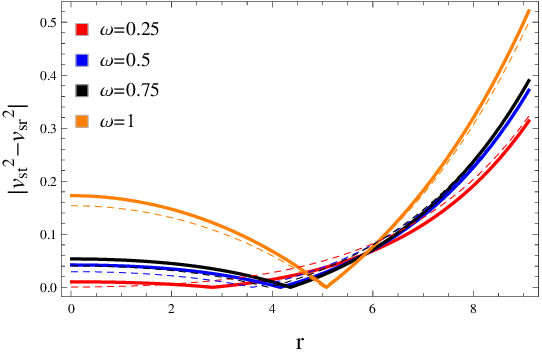,width=0.4\linewidth}\epsfig{file=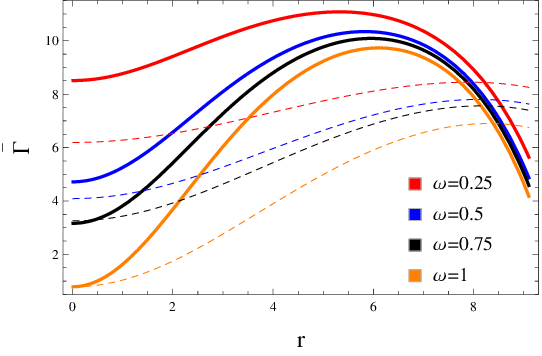,width=0.4\linewidth}
\caption{Radial/tangential speeds of sound, $|v_{st}^2-v_{sr}^2|$
and adiabatic index corresponding to $\mathcal{Q}=0.1$ (solid) and
$0.8$ (dashed) for the solution generated by
$\bar{\mathrm{Y}}_{TF}=0$.}
\end{figure}

\section{Conclusions}

This paper investigates the nature of two different anisotropic
spherically symmetric solutions under the impact of an
electromagnetic field through a systematic scheme, referred as the
gravitational decoupling. For this, we have considered a static
sphere configured with the anisotropic interior as a seed source and
added another fluid distribution. We have then developed the field
equations possessing both matter sources, however, it was hard
enough to find their solution due to the increment in unknowns. This
problem was handled by applying the $\mathrm{MGD}$ technique that
divides these equations into two independent sets. We have observed
that charge contributes only in the first set characterizing the
initial fluid source. Since we need to find multiple solutions, thus
we have adopted some specific metric ansatz as
$$\xi_1(r)=\ln\bigg\{\mathrm{C}_2^2\bigg(1+\frac{r^2}{\mathrm{C}_1^2}\bigg)\bigg\},\quad
\xi_4(r)=e^{-\xi_2(r)}=\frac{\mathrm{C}_1^2+r^2}{\mathrm{C}_1^2+3r^2},$$
and Tolman IV components to deal with the first set. These
spacetimes contain an unknown triplet
($\mathrm{C}_1,\mathrm{C}_2,\mathrm{C}_3$) that needed to be
calculated. To do this, we have taken the Reissner-Nordstr\"{o}m
metric and matched it with the interior geometry at the spherical
junction that made the above triplet known. As for the other set
\eqref{g21}-\eqref{g23} is concerned, we have assumed different
constraints on the additional field $\mathrm{Z}_{\lambda\chi}$ to
determine the deformation function, ultimately resulting in
different solutions. The first model was based on the assumption
that we can convert the anisotropic system to an isotropic interior
for a specific parametric value, i.e., $\omega=1$. Moreover, the
total fluid configuration was reviewed to be free from complexity,
leading to the second model.

We have explored the graphical nature of the developed models by
adopting multiple values of charge and the decoupling parameter to
observe how these models behave in the considered scenario. Further,
we have assumed an interior charge in a particular form involving a
constant whose value has been suggested as $\xi_5=-0.001$
\cite{42aaa} to get acceptable results. We have adopted different
values of this constant and deduced that only the suggested value
provided physically relevant properties of a compact star. In
addition, we have calculated the values of $\mathrm{D}_1$ and
$\mathrm{D}_3$ by taking  into account the vanishing radial pressure
at the spherical boundary. Different parameters such as the
deformation function, extended radial metric component, the matter
triplet \big(defined in \eqref{g46}-\eqref{g48} and
\eqref{60g}-\eqref{60i}\big), anisotropic factors, the mass
function, redshift, compactness and the viability conditions have
been checked for both solutions and found an acceptable profile. We
have also observed that the resulting anisotropic interior
corresponding to $\bar{\mathrm{Y}}_{TF}=0$ becomes less dense for
every $\omega$ as compared to the other model. Figures \textbf{5}
and \textbf{13} show that both the developed solutions fulfil the
hydrostatic equilibrium condition for every parametric choice.

The stability of these structures has also been studied through the
sound speed and adiabatic index so that we can check whether the
$\mathrm{MGD}$ strategy on the charged interiors results in
acceptable results or not. Both these criteria revealed that our
resulting solutions are stable everywhere for all values of $\omega$
and charge except $\omega=1$, contradicting the uncharged framework
\cite{37k} as well as Brans-Dicke theory \cite{37m} (Figures
\textbf{6} and \textbf{14}). Sharif and Sadiq \cite{35} formulated
two new decoupled charged anisotropic models through two
constraints. They analyzed the effects of both decoupling parameter
and an electromagnetic field on them, and found them unstable for
the considered parametric values. Hence, we can say that the
$\mathrm{MGD}$ approach along with the complexity of a compact model
produce more efficient results. To compare our results with the
observational data, we have adopted a compact star $4U~1820-30$
along with its estimated radius and mass. Tables
\textbf{1}-\textbf{4} portray the numerical values of the central
and surface density for all chosen values of the decoupling
parameter and charge. We have found them of order $10^{14}$ or
$10^{15}$, which is sufficiently high and consistent with compact
stars. We have also calculated the mass of a considered compact star
as
\begin{itemize}
\item $\mathcal{M}=1.531\mathcal{M}_{\bigodot}$ and $1.499\mathcal{M}_{\bigodot}$ for $\mathcal{Q}=0.1$ and $0.8$, respectively, corresponding to the model generated by $\bar{\Pi}=0$,
\item $\mathcal{M}=1.538\mathcal{M}_{\bigodot}$ and $1.501\mathcal{M}_{\bigodot}$ for $\mathcal{Q}=0.1$ and $0.8$, respectively, corresponding to the model generated by $\bar{\mathrm{Y}}_{TF}=0$.
\end{itemize}
Hence, it is observed that the lower values of charge produce a best
fit to the existing data. It is worth mentioning that our results
reduce to \cite{37k} for the vanishing charge.\\
\textbf{Data Availability Statement:} This manuscript has no
associated data.


\begin{thebibliography}{43}
\bibitem{1} Schwarzschild, K.: Sitz. Deut. Akad. Wiss Berlin Kl. Math. Phys. \textbf{1916}(1916)189.

\bibitem{2} Schwarzschild, K.: Sitz. Deut. Akad. Wiss Berlin Kl. Math. Phys. \textbf{24}(1916)424.

\bibitem{29} Ovalle, J.: Mod. Phys. Lett. A \textbf{23}(2008)3247.

\bibitem{30} Ovalle, J. and Linares, F.: Phys. Rev. D \textbf{88}(2013)104026.

\bibitem{3} Ovalle, J.: Int. J. Mod. Phys. D \textbf{18}(2009)837.

\bibitem{4} Ovalle, J.: Mod. Phys. Lett. A \textbf{25}(2010)3323.

\bibitem{4a} Sharif, M. and Naseer, T.: Chin. J. Phys. \textbf{73}(2021)179; Eur. Phys. J. Plus
\textbf{137}(2022)1304; Gen. Relativ. Gravit. \textbf{55}(2023)87; Naseer, T. and Sharif, M.: Universe
\textbf{8}(2022)62; Fortschr. Phys.
\textbf{71}(2023)2300004.

\bibitem{5} Casadio, R. and Ovalle, J.: Phys. Lett. B \textbf{715}(2012)251.

\bibitem{6} Casadio, R. and Ovalle, J.: Gen. Rel. Grav. \textbf{46}(2014)1669

\bibitem{7} Ovalle, J., Linares, F., Pasqua, A. and Sotomayor, A.: Class. Quantum Grav.
\textbf{30}(2013)175019.

\bibitem{8} Casadio, R. and Harms, B.: Phys. Rev. D \textbf{64}(2001)024016.

\bibitem{9} da Rocha, R. and Hoff da Silva, J.M.: Phys. Rev. D
\textbf{85}(2012)046009.

\bibitem{10} Bazeia, D., Hoff da SIlva, J.M. and da Rocha, R.: Phys. Lett. B
\textbf{721}(2013)306.

\bibitem{11} da Rocha, R., Piloyan, A. and Kuerten, A.M.: Class. Quantum Grav.
\textbf{30}(2013)045014.

\bibitem{12} Herrera-Aguilar, A., Kuerten, A.M. and da Rocha, R.: Adv. High
Energy Phys. \textbf{2015}(2015)359268.

\bibitem{31} Casadio, R., Ovalle, J. and Da Rocha, R.: Class. Quantum Grav. \textbf{32}(2015)215020.

\bibitem{13} Jeans, J.: Mon. Not. R. Astron. Soc. \textbf{82}(1922)122.

\bibitem{14} Ruderman, M.: Annu. Rev. Astron. Astrophys. \textbf{10}(1972)427.

\bibitem{15} Yazadjiev, S.S.: Phys. Rev. D \textbf{85}(2012)044030.

\bibitem{16} Cardall, C.Y., Prakash, M. and Lattimer, J.M.: Astrophys. J. \textbf{554}(2001)322.

\bibitem{17} Ciolfi, R., Ferrari, V. and Gualtieri, L.: Mon. Not. R. Astron. Soc. \textbf{406}(2010)2540.

\bibitem{18} Frieben, J. and Rezzolla, L.: Mon. Not. R. Astron. Soc. \textbf{427}(2012)3406.

\bibitem{19} Sawyer, R.F.: Phys. Rev. Lett. \textbf{29}(1972)382.

\bibitem{20} Canuto, V. Annu. Rev. Astron. Astrophys. \textbf{12}(1974)167.

\bibitem{21} Heiselberg, H. and Hjorth-Jensen, M.: Phys. Rep. \textbf{328}(2000)237.

\bibitem{33} Ovalle, J. et al.: Eur. Phys. J. C \textbf{78}(2018)960.

\bibitem{36a} Estrada, M. and Tello-Ortiz, F.: Eur. Phys. J. Plus \textbf{133}(2018)453.

\bibitem{36} Gabbanelli, L., Rinc{\'o}n, {\'A}. and Rubio, C.: Eur. Phys. J. C \textbf{78}(2018)370.

\bibitem{37a} Hensh, S. and Stuchl{\'\i}k, Z.: Eur. Phys. J. C \textbf{79}(2019)834.

\bibitem{35} Sharif, M. and Sadiq, S.: Eur. Phys. J. C \textbf{78}(2018)410.

\bibitem{35a} Sharif, M. and Saba, S.: Eur. Phys. J. C \textbf{78}(2018)921;
Sharif, M. and Waseem, A.: Ann. Phys. \textbf{405}(2019)14; Sharif,
M. and Majid, A.: Chin. J. Phys. \textbf{68}(2020)406.

\bibitem{37f} Sharif, M. and Naseer, T.: Phys. Scr. \textbf{97}(2022)055004; ibid. \textbf{97}(2022)125016;
Int. J. Mod. Phys. D \textbf{31}(2022)2240017; Fortschr. Phys.
\textbf{71}(2023)2200147.

\bibitem{37fa} Bekenstein, J.D.: Phys. Rev. D \textbf{4}(1971)2185.

\bibitem{37fb} Esculpi, M. and Aloma, E.: Eur. Phys. J. C \textbf{67}(2010)521.

\bibitem{42aaa} de Felice, F., Yu, Y.Q. and Fang, J.: Mon. Not. R. Astron. Soc.
\textbf{277}(1995)L17.

\bibitem{37fc} de Felice, F., Liu, S.M. and Yu, Y.Q.: Class. Quantum
Gravit. \textbf{16}(1999)2669.

\bibitem{37fd} Maurya, S.K. et al.: Eur. Phys. J. C \textbf{75}(2015)389.

\bibitem{37g} Herrera, L.: Phys. Rev. D \textbf{97}(2018)044010.

\bibitem{37h} Herrera, L., Di Prisco, A. and Ospino, J.: Phys. Rev. D
\textbf{98}(2018)104059.

\bibitem{37i} Yousaf, Z., Bhatti, M.Z. and Naseer, T.: Eur. Phys. J. Plus
\textbf{135}(2020)353; Phys. Dark Universe \textbf{28}(2020)100535;
Int. J. Mod. Phys. D \textbf{29}(2020)2050061; Ann. Phys.
\textbf{420}(2020)168267.

\bibitem{37ia} Yousaf, Z. et al.: Phys. Dark Universe
\textbf{29}(2020)100581; Yousaf, Z. et al.: Mon. Not. R. Astron.
Soc. \textbf{495}(2020)4334; Sharif, M. and Naseer, T.: Chin. J.
Phys. \textbf{77}(2022)2655; Eur. Phys. J. Plus
\textbf{137}(2022)947.

\bibitem{37j} Carrasco-Hidalgo, M. and Contreras, E.: Eur. Phys. J. C
\textbf{81}(2021)757; Andrade, J. and Contreras, E.: Eur. Phys. J. C
\textbf{81}(2021)889.

\bibitem{37k} Casadio, R. et al.: Eur. Phys. J. C
\textbf{79}(2019)826.

\bibitem{37m} Sharif, M. and Majid, A.: Eur. Phys. J. Plus
\textbf{137}(2022)114.

\bibitem{37l} Maurya, S.K. and Nag, R.: Eur. Phys. J. C
\textbf{82}(2022)48; Maurya, S.K. et al.: Eur. Phys. J. C
\textbf{82}(2022)100.

\bibitem{37la} Arias, C. et al.: Ann. Phys.
\textbf{436}(2022)168671; Sharif, M. and Naseer, T.: Ann. Phys.
\textbf{453}(2023)169311.

\bibitem{39} Sharif, M. and Naseer, T.: Class. Quantum Gravit.
\textbf{40}(2023)035009.

\bibitem{42a1} Einstein, A.: Ann. Math. \textbf{40}(1939)922.

\bibitem{42aa} G{\"u}ver, T., Wroblewski, P., Camarota, L. and
{\"O}zel, F.: Astrophys. J. \textbf{719}(2010)1807.

\bibitem{42a} Buchdahl, H.A.: Phys. Rev. \textbf{116}(1959)1027.

\bibitem{42b} Ivanov, B.V.: Phys. Rev. D \textbf{65}(2002)104011.

\bibitem{42bb} Abreu, H., Hernandez, H. and Nunez, L.A.: Class. Quantum Gravit.
\textbf{24}(2007)4631.

\bibitem{42ba} Herrera, L.: Phys. Lett. A \textbf{165}(1992)206.

\bibitem{44} Heintzmann, H. and Hillebrandt, W.: Astron. Astrophys.
\textbf{38}(1975)51.
\end{thebibliography}
\end{document}